\documentclass[letterpaper,aps,showpacs,twocolumn,eqsecnum,superscriptaddress]{revtex4}

\usepackage{amssymb}
\usepackage{amsmath}
\usepackage{latexsym}
\usepackage{epsfig}
\usepackage{longtable}

\newcommand{\be}{\begin{equation}}
\newcommand{\ee}{\end{equation}}
\newcommand{\bea}{\begin{eqnarray}}
\newcommand{\eea}{\end{eqnarray}}

\begin{document}

%%%%%%%%%%%%%%%%%
%%%   TITLE   %%%
%%%%%%%%%%%%%%%%%

\title{The Einstein-Maxwell system in 3+1 form and initial data for
multiple charged black holes}

\author{Miguel Alcubierre}
\email{malcubi@nucleares.unam.mx}
\affiliation{Instituto de Ciencias Nucleares, Universidad Nacional
Aut\'onoma de M\'exico, A.P. 70-543, M\'exico D.F. 04510, M\'exico.}

\author{Juan Carlos Degollado}
\email{jcdegollado@nucleares.unam.mx}
\affiliation{Instituto de Ciencias Nucleares, Universidad Nacional
Aut\'onoma de M\'exico, A.P. 70-543, M\'exico D.F. 04510, M\'exico.}

\author{Marcelo Salgado}
\email{marcelo@nucleares.unam.mx}
\affiliation{Instituto de Ciencias Nucleares, Universidad Nacional
Aut\'onoma de M\'exico, A.P. 70-543, M\'exico D.F. 04510, M\'exico.}

%%%%%%%%%%%%%%%%
%%%   DATE   %%%
%%%%%%%%%%%%%%%%

\date{\today}

%%%%%%%%%%%%%%%%%%%%
%%%   ABSTRACT   %%%
%%%%%%%%%%%%%%%%%%%%

\begin{abstract}
We consider the Einstein-Maxwell system as a Cauchy initial value
problem taking the electric and magnetic fields as independent
variables. Maxwell's equations in curved spacetimes are derived in
detail using a 3+1 formalism and their hyperbolic properties are
analyzed, showing that the resulting system is symmetric
hyperbolic. We also focus on the problem of finding initial data for
multiple charged black holes assuming time-symmetric initial data and
using a puncture-like method to solve the Hamiltonian and the Gauss 
constraints. We study the behavior of the resulting initial data families, and show
that previous results in this direction can be obtained as particular
cases of our approach.
\end{abstract}

%%%%%%%%%%%%%%%%
%%%   PACS   %%%
%%%%%%%%%%%%%%%%


\maketitle

%%%%%%%%%%%%%%%%%%%%%%%%
%%%   INTRODUCTION   %%%
%%%%%%%%%%%%%%%%%%%%%%%%

\section{Introduction}
\label{sec:introduction}

One of the most important results in numerical relativity in recent
years has been the successful simulation of the coalescence of two
spiraling (spin or spin-less) black holes (see
Refs.~\cite{Pretorius:2009nq,Hannam:2009rd} for an overview). This
simulations are important since one expects black hole collisions to
be among the most powerful sources of gravitational radiation.
Gravitational radiation from this type of sources will presumably be
measured by the next generation of interferometric gravitational
observatories within the next decade or so~\cite{Aylott:2009ya}. It is
therefore important to have simulations of different kind of
astrophysical scenarios in order to compare with the observational
results in order to reach a deeper understanding of such sources of
gravitational radiation.  In addition to the simulations of coalescing
black holes, different authors have started to analyze collisions of
extended objects ({\em i.e.} objects constructed with a non-zero
energy momentum tensor) like neutron
stars~\cite{Kiuchi:2009jt,Giacomazzo:2009mp,Anderson2008b,Anderson2008a,Baiotti:2008ra}
or even more
exotic objects like boson
stars~\cite{Palenzuela:2007dm,Palenzuela:2006wp}. Since such objects
are less compact than black holes, and since their individual masses
are limited, one expects that the amount of gravitational radiation
emitted by the collision of these objects will be weaker relative to
the two-black hole problem. However, from the numerical point of view,
such scenarios are by far more challenging since hydrodynamics,
micro-physics or field theory are also involved. Moreover, their
gravitational-wave signals can be richer in the sense that they can
carry information about the internal composition ({\em e.g.} equation
of state).
 
Another interesting numerical problem that one can conceive and that
might have an observational counterpart is the collision of two
charged black holes (TCBH). This problem is simpler than the case of
two neutron stars but perhaps more interesting than the two uncharged
black hole collision.  In fact, it is conceivable that if black holes
in binary systems were formed by the gravitational core collapse of
neutron stars or supernovae then they could have a small amount of
charge. Actually, even if a black hole is originally uncharged but
immersed in a uniform magnetic field (which in turn can be produced by
an accreting plasma surrounding the back hole), it can be charged up
to some extent~\cite{Wald74}. Very likely their charge would be small
compared with their mass (in suitable units), but perhaps large enough
to leave an imprint in the wave forms of gravitational radiation
during a collision. In fact, numerical simulations of electromagnetic
fields immersed in the background spacetime corresponding to the
collision of two uncharged black-holes already show that the dynamics
of the background spacetime induce the emission of electromagnetic
radiation that is correlated in a very particular way with the
gravitational-wave signals~\cite{Palenzuela:2009yr}. Such
electromagnetic radiation (if detected) might provide information
about the pre-merger stage as a precursor to the coalescence of the
black holes.  One can expect that taking into account the
back-reaction of the electromagnetic field in the spacetime itself can
have even more interesting features.

But even from the theoretical point of view, it seems important to
analyze the interplay between the gravitational and electromagnetic
forces in a TCBH collision. Indeed, the analysis of the interplay
between electromagnetism and gravity within a BH spacetime has a long
history in general relativity~\footnote{In a wider context which
  includes extended objects and not only black holes, such an
  interplay was first analyzed by Wheeler
  in~\cite{Wheeler1955}. However, here we will not be concerned with
  objects other than charged black holes.}.  The first analytic
solution to Einstein's equations involving an electromagnetic field
was given by Reissner and
Nordstr\"om~\cite{Reissner16,Nordstrom18}. This solution was
interpreted as a spacetime containing a static and spherically
symmetric charged black hole.  Much more later,
Papapetrou~\cite{Papapetrou1945} and Majumdar~\cite{Majumdar1947}
found a static solution involving multiple black holes having the
maximal charge-to-mass ratios (see Sec. \ref{sec:exactdata}).
Perjes~\cite{Perjes1971}, and Israel and Wilson~\cite{Israel1972}
generalized this solution to the stationary case.  Several uniqueness
theorems within the Einstein-Maxwell theory have been established
during the past which show how to characterize certain kind of
stationary black hole solutions (see Ref.~\cite{Heusler1996} for a
review).

More recently, Ruffini and collaborators~\cite{Bini2007} have analyzed
the case of a static charge in a Reissner-Nordstr\"om spacetime by
using a perturbative approach.  In the framework of the so-called {\em
  membrane paradigm}~\cite{Thorne86}, and using a 3+1 formalism,
Thorne and colleagues have dealt with different problems involving
electromagnetic fields in strong gravitational background fields
(namely those generated by stationary black holes). Such a
``membrane'' viewpoint allows one to approximate several results
concerning electromagnetic fields around BH's, notably at the
horizons. The membrane viewpoint assigns to the horizon thermodynamic,
mechanical and electric properties. In this direction one can mention
the pioneering work of Damour~\cite{Damour78} and
Znajek~\cite{Znajek78}, who analyzed the boundary conditions of
electromagnetic fields at the horizons of a BH. Such boundary
conditions can be thought of as arising from the physical properties
of a fictitious membrane residing at the horizon or near the horizon
(see~\cite{Thorne86} for the introduction of the notion of ``stretched
horizon'' which allows to approximate these boundary conditions near
the true horizon).  Since in this viewpoint one neglects the
back-reaction of the matter fields into the spacetime, this
approximation will break down in situations where the self-gravity of
the matter is important. It is in this regime where numerical
relativity becomes crucial.

In the context of multiple charged black holes, the work by
Bowen~\cite{Bowen85} was one of the first to address the initial data
problem. He considered the case of zero initial magnetic field, but
without imposing a moment of time-symmetry and without resorting to
electromagnetic potentials.  We will often make reference to Bowen's
work in this paper.  Previous to Bowen's work,
Lindquist~\cite{Lindquist1963} generated initial data for many
stationary charged particles by imposing time-symmetry and using a
method of images and electromagnetic potentials.  The present paper is
similar in spirit to Bowen's work, except that we shall mainly be
interested in a TCBH initial data which is computed using a method
analogous to the puncture method~\cite{Brandt97} which is not
inversion symmetric (see Sec. \ref{sec:initial}).

In order to tackle the problem of a TCBH collision there are many
challenges, both numerical and analytical. First, from the numerical
point of view there are two main considerations: one is the initial data and
another one is the evolution of the Einstein-Maxwell system. For the
former one requires suitable initial data compatible with the
constraint equations. Whereas for the latter one needs a numerical
code to solve the Einstein-Maxwell evolution system together with all
the numerical tools necessary to analyze the location of horizons, the
amount of emission of gravitational and electromagnetic radiation,
etc.

As regards the analytical challenges, there are various issues with
different levels of complexity. Since the vast majority of numerical
relativists are concerned with the 3+1 formulation of Einstein
equations and their corresponding numerical solution, the first step
toward our goal consists in obtaining a well defined 3+1 decomposition
of Maxwell's equations in a curved spacetime. It turns out that this
problem has been embraced in the past in at least two works: first by
Ellis in~\cite{Ellis73}, and later by Thorne \& Macdonald
in~\cite{Thorne82}. In Section~\ref{sec:Maxwell} we present a
completely independent derivation of the 3+1 Maxwell equations which
we then compare with the one reported by Thorne \& Macdonald. Our
derivation is based primary in the 3+1 formulation considered by
York~\cite{York79} (see Refs.~\cite{Alcubierre08a,Gourgoulhon07} for a
thorough review of this formulation).  As we will show, the 3+1
Maxwell equations in curved spacetime are in fact very similar to the
usual Maxwell equations in flat spacetime, except for the fact that
some extra terms due to the curvature arise. We obtain two ``scalar''
constraint equations for the electric and magnetic fields, and two
``vector'' evolution equations for those fields. In this case, the
electric and magnetic fields are referred to the so-called {\em
  Eulerian} observers whose four-velocity is orthogonal to the
spacelike hypersurfaces that define the foliation of the four
dimensional spacetime (see Sec. \ref{sec:Maxwell}). The advantage of
this 3+1 formulation of Maxwell equations, as opposed to the one based
on electromagnetic potentials, is that their fundamental variables are
gauge invariant {\em ab-initio}. Therefore, one needs to focus only in
the gravitational gauge issue. Moreover, the set of equations turns
out to be manifestly hyperbolic (symmetric hyperbolic). The only
difference with respect to the flat case is that the eigenvalues and
eigenvectors associated with the principal part of the equations
include contributions due to the (densitized) lapse and shift. We
analyze these aspects in Sec.~\ref{sec:hyperbolicity}.

As we mentioned above, another challenge which can be both numerical
and analytical concerns the initial data. Let us recall that in the
case of the two uncharged BH collision one can consider a rich family
of interesting initial conditions. The simpler ones consist on
imposing a moment of time symmetry in which the two BH are initially
at rest and without angular momentum and spin. With such a condition
the momentum constraints are satisfied trivially. In this case, the
initial data will generate a head-on collision. Moreover, one can
assume that the 3-metric is conformally flat. Both conditions in the
Hamiltonian constraint lead to an elliptic equation for the conformal
factor (see Sec. \ref{sec:conformal}).  A unique solution of that
equation is obtained when imposing suitable boundary conditions which
in turn are related to the topology of the initial hypersurface
$\Sigma_0$ (see Refs.~\cite{Alcubierre08a,Gourgoulhon07b} for a
pedagogical review on initial data).  One possible topology for
$\Sigma_0$ is $\mathbb{R}^3$ minus two balls (the boundary of which
represent the horizons of the two BH). This kind of initial data has
been used in the so-called excised approach where one ignores what is
inside those balls while evolving the rest of the spacetime. A simpler
topology for $\Sigma_0$ is to consider $\mathbb{R}^3$ minus two
``points''. This is the so-called ``puncture data'' approach. The
punctures represent in fact two different asymptotically flat regions
on a two-throat shaped spacelike $\Sigma_0$. Since in vacuum the
elliptic equation for the conformal factor is indeed linear one can in
fact construct puncture data that represent an arbitrary number of
black holes that are initially at rest. This data has a simple
analytical expression whose form resemble the electric potential
generated by a series of point charges.

A much more realistic initial data was the one used in the simulations
of the coalescence of two binary BH that we alluded above. In this
case, there is a remarkable analytical solution (the Bowen-York
solution) of the momentum constraints which represents two BH with
arbitrary linear momentum and spin~\cite{Bowen80}. This solution is
constructed using a conformal and transverse-traceless
decomposition~\cite{Gourgoulhon07b}.  The most difficult part of this
approach is to solve a highly non-linear elliptic equation for the
conformal factor. Again, one can adopt the puncture or the excised
approach. In either case one needs to solve the elliptic equation
numerically. The puncture approach has been very popular recently and
is the one that has led to many successful evolutions of different BH
binary configurations. The outcome of such evolutions has led to the
prediction of a large variety of gravitational wave forms that in the
near future will be confronted with the observational
data~\cite{Aylott:2009ya}.

Returning now to the case of the two-charged BH problem, one can take
advantage of the experience with the uncharged case in order to
construct interesting initial data. Unlike the uncharged case, the new
difficulty is that the Einstein constraint equations will have
contributions due to the electromagnetic fields. In order to simplify
the problem the first attempt consists in assuming again a moment of
time symmetry where the magnetic field is zero. As mentioned before,
this kind of initial data was considered in the past by
Bowen~\cite{Bowen85}. In this case the Poynting vector (which is the
source of the momentum constraints) becomes identically zero. We could
then use the Bowen-York initial data for this problem. However, one
still needs to solve the Hamiltonian constraint plus the Gauss
constraint for the electric field, the former containing now the
contribution due to the electrostatic energy-density associated with
the initial electric field generated by the two-charged BH. One can
further simplify the problem if one assumes that the two BH are
initially at rest (zero spin and zero linear and angular
momentum). This initial data would then represent the head-on
collision of two charged BH. However, unlike the vacuum case, the
elliptic equation for the conformal factor is now highly non-linear.
Remarkably, we have found a way to solve analytically both constraints
(the Hamiltonian and Gauss constraints) using a puncture approach (see
Sec.~\ref{sec:exactdata}). This solution represents a superposition of
multiple charged black holes all of which having the same
charge-to-mass ratio. When this last condition is dropped, finding an
analytical solution seems difficult. However, it is not difficult to
find numerical solutions (see Sec.~\ref{sec:numericaldata}).

The paper is organized as follows: Section~\ref{sec:Maxwell} presents
our derivation of the 3+1 Maxwell equations in a curved spacetime, as
well as the analysis of their hyperbolic properties (this section is
complemented by an Appendix). The Einstein-Maxwell system is summarized 
and a brief discussion on the electromagnetic 
potentials is also included. In Section~\ref{sec:initial} we analyze
the initial data for multiple black holes using a conformal approach,
where both analytical and numerical results are obtained. Finally
Section~\ref{sec:conclusions} contains several comments and remarks
for the future.

%%%%%%%%%%%%%%%%%%%%%%%%%%%%%%%%%%%%%%%%%
%%%   MAXWELL EQUATIONS IN 3+1 FORM   %%%
%%%%%%%%%%%%%%%%%%%%%%%%%%%%%%%%%%%%%%%%%

\section{The Maxwell equations in 3+1 form}
\label{sec:Maxwell}

In the following we assume that the reader is familiar with the 3+1
formalism of general relativity (see
Refs.~\cite{York79,Gourgoulhon07,Alcubierre08a}, for a thorough
review, and~\cite{Wald84} for the conventions adopted here).

Let us first remember that the covariant Maxwell equations read
\begin{eqnarray}
\label{Maxcov1}
\nabla_a F^{ab} &=& -4 \pi j^b \; , \\
\label{Maxcov2}
\nabla_a F^{*ab} &=& 0 \; ,
\end{eqnarray}
where 
\begin{eqnarray}
\label{Faraday}
F_{ab} &=& \partial_a A_b - \partial_b A_a \; , \\
\label{FaradayDual}
F^{*ab} &:=& -\frac{1}{2} \: \epsilon^{abcd}F_{cd} \; ,
\end{eqnarray}
are the Faraday's tensor and its dual. Here we take the convention
that $\epsilon^{0123}= - 1/\sqrt{-g}$ and $\epsilon_{0123}=
+\sqrt{-g}$, with the signature of $g_{ab}$ taken as $(-\,,\,+\,,\,+\,,\,+)$
\footnote{A change of convention for the signs of the Levi-Civita
  tensor (for a left-handed as opposed to right-handed orthonormal
  basis), entails a change of sign in equations~\eqref{FaradayDual}
  and~\eqref{Magnetic1}. For instance, the convention adopted in many
  books like Landau's {\em The Classical Theory of Fields} is that
  $\epsilon_{0123}= - 1$ and $\epsilon^{0123}= + 1$ in a left-handed
  orthonormal basis.} .

In order to obtain the 3+1 decomposition of Maxwell equations from
their covariant form, on has to proceed in a way very similar to the
derivation of the so called Arnowitt-Deser-Misner (ADM) equations of
General Relativity. Let us briefly review the ingredients of this
decomposition.  One considers a spacetime $(M,g_{ab})$ (assumed to be
globally hyperbolic) which is foliated by a family of spacelike
hypersurfaces $\Sigma_t$ ($t\in \mathbb{R}$) parametrized by a global
time function $t$ ({\em i.e.} $M$ has topology \mbox{$M=\Sigma_t
  \times \mathbb{R}$}). The foliation is achieved in the following
way: On a Cauchy surface $\Sigma_t$, one is given an initial data set
that satisfies some constraint equations (see
Sec.~\ref{sec:EinsteinMaxwell}).  The full spacetime is
``reconstructed'' by evolving these initial data using a suitable set
of evolution equations (which includes the gauge). Such a set of
constraints and evolution equations are known as the ADM equations of
General Relativity.

In order to find a similar set of equations for the electromagnetic
case, an analogous algebraic and geometric decomposition of the field
equations~\eqref{Maxcov1} and~\eqref{Maxcov2} has to be performed.
The general procedure for obtaining a 3+1 splitting of a system of
covariant field equations (also called {\em orthogonal decomposition})
consists on projecting the different tensor fields in the directions
parallel and orthogonal to the timelike unit vector field $n^a$ ($n^a
n_a=-1$) which is normal to $\Sigma_t$.  The projection onto
$\Sigma_t$ is performed by first defining the {\em projector operator}:
\begin{equation}
\label{proj}
h^a_{\,\,b} = \delta^a_{\,\,b} + n^a n_b \; .
\end{equation}
This tensor field has the property of being idempotent:
\mbox{$h_a^{\,\,c}h_c^{\,\,b}= h_a^{\,\,b}$}.

A tensor field $\,\!^3T^{a_1 a_2 ... a_k}_{\hspace{1.1cm}b_1
  b_2...b_l}$ is said to be tangential to $\Sigma_t$ if, when
contracted with $n^a$ it gives zero, or equivalently if contracted
with $h^a_{\,\,b}$ it remains unchanged. For brevity, such tensors
will be termed $3-$tensors. Any tensor field can be decomposed {\em
  orthogonally}\/ by using $h^a_{\,\,b}$ and $n^a$. In particular, a
4-vector $w^a$ is decomposed as follows
\begin{equation}
\label{ortdecomw}
w^a =\,\!^3w^a + w_\perp n^a \; ,
\end{equation}
where $\,\!^3w^a:= h^a_{\,\,b} w^b$ and $w_\perp:= -n_c
w^c$. Moreover, the 3+1 splitting of the metric reads
\begin{equation}  \label{3+1metric}
ds^2= -(N^2 - N^i N_i) \: dt^2 - 2 N_i \: dt dx^i + h_{ij} \: dx^i dx^j \; ,
\end{equation}
where the {\em lapse}\/ function $N>0$ is defined as to normalize the
(future pointing) dual vector field $n_a= -N \nabla_a t$. The shift
vector is given by $N^a:= -h^a_{\,\,b} t^b $, where $t^a =
\left(\partial/\partial t\right)^a$ is a vector field that represents
the ``flow'' of the time lines and which satisfies $t^a\nabla_a
t=1$. This means that $t^a$ is orthogonally decomposed as $t^a= -N^a +
Nn^a$, with $N=-n_a t^a$.  Here $h_{ij}$ is the 3-metric (or induced
metric) of the manifold $\Sigma_t$.  In order to avoid confusion a
note on notation is important at this point: many numerical relativity
references (see {\em e.g.}  \cite{Alcubierre08a,York79}) use $\alpha$
and $\beta^a$ to denote the lapse and shift instead of $N$ and $N^a$,
with $\alpha = N$ and $\beta^a = -N^a$ (do note the change in sign in
the shift!).

Another important object is the extrinsic curvature of the embeddings
$\Sigma_t$ which is defined as\footnote{It is to note that in Wald's book 
({\em op. cit.}) the extrinsic curvature is defined with the opposite sign.}
\begin{equation} 
\label{K_ab}
K_{ab}:= -\frac{1}{2} \: {\cal L}_{\mathbf{n}} h_{ab} \; ,
\end{equation}
where ${\cal L}_{\mathbf{n}}$ stands for the Lie derivative
along $n^a$. From the above definition one can obtain the following
identity
\begin{equation}
\label{K_abexp}
K_{ab}= - h^{\,\,c}_{a}  h^{\,\,d}_{b} \nabla_c n_d \; ,
\end{equation}
which shows that $K_{ab}$ is in fact a 3-tensor field. Furthermore,
its trace is given by
\begin{equation}
\label{TraceK}
K = - \nabla_c n^c \; .
\end{equation}

As is well known, the set $(\Sigma_t, h_{ab},K_{ab})$ provides the
initial data for the gravitational field.  This data in fact cannot be
chosen freely, and has to satisfy the Einstein constraint equations
(see Sec.~\ref{sec:EinsteinMaxwell}).

At this point it is useful to introduce a covariant derivative
operator compatible with $h_{ab}$. Given a 3-tensor field
$\,\!^3T^{a_1 a_2 ...a_k}_{\hspace{1.1cm}b_1 b_2 ... b_l}$, one
defines
\begin{eqnarray}
\label{DT}
&& \hspace{-6mm} D_e \,\!^3T^{a_1 a_2 ... a_k}_{\hspace{1.1cm} b_1 b_2 ... b_l}
= \nonumber \\
&& h^{a_1}_{\,\,c_1} ... h^{a_k} _{\,\,c_k} 
h^{\,\,d_1}_{b_1} ...h^{\,\,d_l}_{b_l} h^{f}_{e} \nabla_f \,
\!^3T^{c_1 c_2...c_k}_{\hspace{1.1cm}d_1 d_2 ... d_l} \; , \hspace{8mm}
\end{eqnarray}
where $D_a h_{bc} \equiv 0$. Finally, one should mention that the
indices of 3-tensors can be raised and lowered with $h^{ab}$ and
$h_{ab}$, and also their contravariant time components are identically
zero. Moreover the projector $h^a_{\,\,b}$ applied to any 3-tensor
field acts as a $\delta^a_{\,\,b}$.
\bigskip

The Maxwell equations written in the form of an initial-data (Cauchy)
problem can be obtained by projecting Eqs.~\eqref{Maxcov1}
and~\eqref{Maxcov2} orthogonally and tangentially to the space-like
hypersurface $\Sigma_t$ using $n^a$ and the projector
operator~\eqref{proj}, respectively.  Note that with respect to a
local coordinate basis adapted to the foliation we have \mbox{$n_a=
  (-N,0,0,0)$} and \mbox{$n^a= (1/N,N^i/N)$}. In particular, in a flat
spacetime one has $n_a= (-1,0,0,0)$ and $n^a= (1,0,0,0)$.

We will need to define suitable quantities associated with the
electromagnetic (EM) field before proceeding with the 3+1
decomposition of the Maxwell equations.  In order to do this, consider
first the 3+1 decomposition of an arbitrary 2-0 tensor, say $H^{ab}$,
which is given as follows
\begin{eqnarray}
\label{Hab1}
 H^{ab} &=& \, \!^{(3)}H^{ab} +  n^a \,^{(3)}\,\! H^{\perp b} \nonumber \\
&& + \,^{(3)}\,\! H^{a \perp} n^b + H^{\perp\perp}n^a n^b \; ,
\end{eqnarray}
with
\begin{eqnarray}
\label{3H}
^{(3)}\,\! H^{ab} &:=& h_{\,\,\,c}^{a} h_{\,\,\,d}^{b} H^{cd} \; ,\\ 
^{(3)}\,\! H^{\perp b} &:=&  -n_c h_{\,\,\,d}^{b} H^{cd} \; ,\\ 
^{(3)}\,\! H^{a \perp} &:=& -n_c h_{\,\,\,d}^{a} H^{dc} \; ,\\ 
\label{Hab5}
H^{\perp\perp} &:=& n_a n_b H^{ab} \; .
\end{eqnarray}

For the particular case where $ H^{ab}= F^{ab}$, the antisymmetry of $
F^{ab}$ implies that $F^{\perp\perp} \equiv 0$, so~\footnote{One could
  define $^{(3)}\,\! F_{ab}:= D_a\,^{(3)} \!A_b - D_b\,^{(3)} \!A_a$
  instead of $^{(3)}\,\!  F_{ab} := F_{cd} \: h_a^{\,\,\,c}
  h_b^{\,\,\,d}$ [cf. Eq.(\ref{3H})]. Fortunately, a straightforward
  calculation using the definition of the covariant derivative $D$ in
  terms of $\nabla$, as well as the definition of 3-vectors in terms
  of 4-vectors, shows that both definitions coincide.  Thus, we do not
  need to introduce different notation for both expressions.  This
  coincidence is due to the skew symmetry of the Faraday's tensor.  An
  inconsistency would arise, however, if one where to define the
  3-Riemann tensor as in Eq.~\eqref{3H}. We know that the projection
  of the 4-Riemann tensor onto $\Sigma_t$ as in Eq.~\eqref{3H} is in
  fact related to the 3-Riemann tensor through one of the
  Gauss-Codazzi equations.  Had one used the same notation for both
  3-tensors, or identified one with the other, a serious inconsistency
  would arise.  Therefore, special care has to be taken when using the
  notation with `$^{(3)}\,\!$' within a projected tensor which arises
  originally from covariant derivatives of 4-tensor fields.  The point
  is that the spatial components of 4-tensors do not necessarily
  coincide with the spatial components of the corresponding
  3-tensors. In the case of the 4-Riemann tensor it is obvious that
  its spatial components involve time components of Christoffel
  symbols and therefore, time components of the metric. Whereas the
  3-Riemann tensor involves the 3-metric solely.  In the case of the
  Faraday's tensor the spatial covariant components coincide in both
  cases.}
\begin{equation}
\label{3+1decomp}
 F^{ab} = \,^{(3)}\,\! F^{ab} +  n^a\, ^{(3)}\,\! F^{\perp b}
+ \,\,^{(3)}\,\! F^{a \perp} n^b   \,\,\,\,.
\end{equation}

We now define the electric and magnetic fields as measured by the {\em
  Eulerian}\/ observers with four velocity $n^a$ in the following
way~\footnote{The Eulerian observers are called also {\em fiducial
    observers}~\cite{Thorne82}, or even ZAMO
  observers~\cite{Bardeen73}, in the case of stationary and
  axisymmetric spacetimes. ZAMO stands for {\em zero angular momentum
    observers}\/ because their angular momentum $L:= n^a\psi_a$ turns
  out to be identically zero. Here $\psi^a= \left(\partial/\partial
  \phi\right)^a$ stands for the rotational Killing vector field which
  generates the isommetries around the axis of symmetry.}
\begin{eqnarray}
\label{Electric}
E^a &:=& -n_b F^{ba} \equiv \, ^{(3)}\,\! F^{\perp a} \; ,\\
\label{Magnetic}
B^a &:=& -n_b F^{*ba} \equiv \,^{(3)}\,\! F^{*\perp a} \; .
\end{eqnarray}
The last equalities in Eqs.~\eqref{Electric} and \eqref{Magnetic}
result by noticing that the above definitions and the antisymmetry of
$ F^{ab}$ and $ F^{*ab}$ both imply that $E^a$ and $B^a$ are in fact
3-vector fields since $n_a E^a\equiv 0 \equiv n_a B^a$, or
equivalently $E^a= h_{\,\,\,d}^{a} E^d$, $B^a= h_{\,\,\,d}^{a}
B^d$. Otherwise such equalities can be corroborated simply by
inserting Eq.~\eqref{3+1decomp} and its corresponding dual in the
definitions above. In this way, one finds that
\begin{equation}
\label{Fdecomp}
 F^{ab} =\, ^{(3)}\,\! F^{ab} + n^a E^b  - E^a n^b  \; .
\end{equation}
On the other hand, from Eqs.~\eqref{Magnetic}, \eqref{FaradayDual} and \eqref{Fdecomp} 
we have
\begin{eqnarray}
B^a &=& \frac{1}{2} n_b \epsilon^{bacd}F_{cd} \nonumber \\
&=& \frac{1}{2} n_b \epsilon^{bacd} \left( ^{(3)}\,\!
F_{cd} + n_c E_d - E_c n_d\right)  \nonumber \\
&=& \frac{1}{2} n_b \epsilon^{bacd}\, ^{(3)}\,\! F_{cd} \; ,
\label{Magnetic1}
\end{eqnarray}
where in the last equality we used the fact that the contraction of
the totally antisymmetric Levi-Civita symbol $\epsilon^{bacd}$ with the symmetric
tensors $n_b n_d$ and $n_b n_c$ vanishes identically.  Moreover
\begin{eqnarray}
B^a &\equiv& h_{\,\,\,e}^{a} B^e= \frac{1}{2} n_b h_{\,\,\,e}^{a}
\epsilon^{becd}\, ^{(3)}\,\! F_{cd} \nonumber \\
&=& \frac{1}{2} n_b
h_{\,\,\,e}^{a} h_{c}^{\,\,\,f} h_{d}^{\,\,\,g} \epsilon^{becd}\,
^{(3)}\,\! F_{fg} \nonumber \\
\label{Magnetic2}
&=& \frac{1}{2} \,^{(3)} \,\!\epsilon^{\perp afg} \,^{(3)}\,\! F_{fg} \; ,
\end{eqnarray}
where
\begin{equation}
\label{3-Levi-Civita1}
^{(3)}\,\!\epsilon^{\perp afg}:=
n_b h_{\,\,\,e}^{a} h_{\,\,\,c}^{f} h_{\,\,\,d}^{g}
\epsilon^{becd} \equiv n_b \epsilon^{bafg} \; .
\end{equation}
The last identity arises when using Eq.~\eqref{proj} in the above
definition, plus the fact that all the contractions between
$\epsilon^{becd}$ and more than one factor $n_a$ vanish
identically~\footnote{It is easy to see that any tensor
  $T_{a_1...a_{q-1}}$ of type $(0,q-1)$ defined as the contraction of
  $n_a$ with a totally antisymmetric tensor $A_{a_1...a_{q}}$ of type
  $(0,q)$ will in turn be a 3-tensor by the same argument. This is
  exactly what happens in Eqs.~\eqref{Electric} and
  \eqref{Magnetic}.}.

Now, since $\epsilon^{0123}= - 1/\sqrt{-g}= - 1/(N\sqrt{h})$, where
\mbox{$h:= {\rm det} (h_{ij})$} and $^{(3)} \,\!\epsilon^{\perp afg}= -N
h_{\,\,\,e}^{a} h_{\,\,\,c}^{f} h_{\,\,\,d}^{g} \epsilon^{0ecd}$, it is
clear that $^{(3)} \,\!\epsilon^{\perp 123}= -N \epsilon^{0123}=
1/\sqrt{h}$. Therefore we simply identify
\begin{equation}
\label{3-Levi-Civita2}
\,^{(3)} \,\!\epsilon^{afg} = \,^{(3)} \,\! \epsilon^{\perp afg}\,\,\,\,,
\end{equation}
with the 3-Levi-Civita symbol defined in such a way that $^{(3)}
\,\!\epsilon^{123}=1/\sqrt{h}$~\footnote{In a similar fashion we can
  define $^{(3)} \,\!\epsilon_{\perp efg}= \,^{(3)}
  \,\!\epsilon_{efg}$ such that $\,^{(3)} \,\!\epsilon_{123}= +
  \sqrt{h}$.  Indeed $\epsilon_{\perp acd}:= n^b \epsilon_{b acd}$, and
  therefore $\epsilon_{\perp 123}= (1/N) \: \epsilon_{0123} = \sqrt{h}$
  (where $\epsilon_{0123}= + \sqrt{-g}= N \sqrt{h}$).}.  It is also
convenient to introduce a ``flat'' Levi-Civita symbol as 
\begin{eqnarray}
\label{Levi-Civita-flatup}
\epsilon^{abc}_F &:=& \sqrt{h} \,^{(3)} \,\!\epsilon^{abc} \; ,\\ 
\label{Levi-Civita-flatdown}
\epsilon_{abc}^F &:=& \frac{1}{\sqrt{h}}\,^{(3)}
\,\!\epsilon_{abc} \; ,
\end{eqnarray}
where $\epsilon^{abc}_F$ and $\epsilon_{abc}^F$ take values $(0,\pm
1)$ as in flat space.

We can also invert Eq.~\eqref{Magnetic2} as follows
\begin{equation}
\label{3FtoB}
\,^{(3)}\,\! F^{ab}= \,^{(3)} \,\!\epsilon^{abc} B_c \; ,
\end{equation}
so that Eq.~\eqref{Fdecomp} now reads~\footnote{Substituting
  Eqs. \eqref{3-Levi-Civita1} and \eqref{3-Levi-Civita2} in
  Eq. \eqref{Fdcomp2}, one obtains $F^{ab} = n_c \epsilon^{abcd} B_d +
  n^a E^b - E^a n^b $, which is in fact the second of Eqs.~(3.2) of
  Thorne \& Macdonald~\cite{Thorne82}.}
\begin{equation}
\label{Fdcomp2}
 F^{ab} = \,^{(3)} \,\!\epsilon^{abc} B_c  + n^a E^b  - E^a n^b \; .
\end{equation}

In the same way we can obtain the following 3+1 decomposition of
$F^{*ab}$:
\begin{equation}
\label{FdcompDual}
 F^{*ab} = -\,^{(3)} \,\!\epsilon^{abc} E_c + n^a B^b - B^a n^b \; .
\end{equation}

We note that, just as in case of flat space, the dual operation maps
the electric and magnetic fields as follows: $E_a \rightarrow B_a$,
$B_a \rightarrow -E_a$~\footnote{In the left-handed convention one has
  $\epsilon^{0123}= +1/\sqrt{-g}$ and $\epsilon_{0123}= -1\sqrt{-g}$.
  One should then take the definition $^{(3)}\,\!\epsilon^{\perp
    afg}:= -n_b \: h_{\,\,\,e}^{a} h_{\,\,\,c}^{f} h_{\,\,\,d}^{g}
  \epsilon^{becd}$, so that $^{(3)}\,\!\epsilon^{\perp efg}$ is always
  positive and therefore independent of the conventions.  Namely,
  $^{(3)}\,\!\epsilon^{\perp 123}= N\epsilon^{0123} = +1/\sqrt{h}$. In
  this way the signs of Eqs. (\ref{Magnetic2}) and
  (\ref{3FtoB})$-$(\ref{FdcompDual}) remain invariant, as well as the
  subsequent derivations which use them, like Maxwell's equations.}.

We are now in the position of performing the 3+1 splitting of Maxwell
equations (details can be found in the Appendix). The projection of
Eqs.~\eqref{Maxcov1}$-$\eqref{Maxcov2} onto $n_a$, after the use of
Eqs.~\eqref{Fdcomp2} and \eqref{FdcompDual}, leads to the initial
value constraints for the electric and magnetic fields respectively:
\begin{eqnarray}
\label{Max3}
D_a E^a &=& 4\pi \rho \; , \\
\label{Max4}
D_a B^a &=& 0 \; .
\end{eqnarray}
where we remind the reader that $D_a$ is the derivative operator
compatible with $h_{ab}$ [cf. Eq.~\eqref{DT}], and \mbox{$\rho:= -n_a
  j^a$} is the charge density as measured by the Eulerian observer.
More specifically, the covariant 3-divergence is given by
\begin{equation}
D_a E^a = \frac{1}{\sqrt{h}} \: \partial_i
\left( \sqrt{h} E^i \right) \; .
\end{equation}

On the other hand, the projection of \eqref{Maxcov1} onto $\Sigma_t$
provides the evolution equation for the electric field (see Appendix
for details):
\begin{eqnarray}
\label{DynE}
h_{\,\,\,c}^{a} \: {\cal L}_{\mathbf{n}} E^c &=& \nonumber \\
&& \hspace{-15mm} (D\times B)^a - (B\times a)^a + K E^a
- 4\pi \,^{(3)}\,\!j^a \; , \hspace{5mm}
\end{eqnarray}
with
\begin{eqnarray}
\label{rotB}
(D\times B)^a &:=& \,^{(3)} \,\!\epsilon^{abc} \partial_b B_c \; ,\\ 
\label{Bxa}
(B\times a)^a &:=& \,^{(3)} \,\!\epsilon^{abc} B_b a_c \; , \\
\,^{(3)}\,\!j^a &:=& h_{\,\,\,b}^{a} j^b \; ,
\end{eqnarray}
and where $K:= K^{a}_{\,\,a}$ is the trace of the extrinsic curvature
given by Eq.~\eqref{TraceK}, and $a^c:= n^a \nabla_a n^c \equiv
D^c({\rm ln} N)$ is the {\em acceleration}\/ of the Eulerian
observer~\cite{York79,Gourgoulhon07}.

Taking the spatial components of Eq.~\eqref{DynE} one
finds~\footnote{This equation can also be written in the following way
  $\label{evE2} \partial_t {\vec E} + {\cal L}_{\mathbf{N}} E^i =
  D_{\rm flat} \times {(\cal N} {\vec B}) + NK {\vec E} - 4\pi
  N\,^{(3)}\,\!{\vec j} $, with ${\cal N}:= N/\sqrt{h}$ the densitized
  lapse, and where the rotational operator is that of flat space (even
  if one is using curvilinear coordinates).  The curvature and
  geometric factors have been absorbed in the densitized lapse.}
\begin{eqnarray}
\partial_t E^i + {\cal L}_{\mathbf{N}} E^i &=&
(D \times N B)^i \nonumber \\
&+& N K E^i - 4 \pi N \: ^{(3)}\!j^i \; ,
\label{evE}
\end{eqnarray}
where now ${\cal L}_{\mathbf{N}}$ is the Lie derivative along the
shift, and where we used the fact that $(D \times N B)^i
= N (D \times B)^i - N (B \times a)^i$.

In a similar way, the projection of \eqref{Maxcov2} onto $\Sigma_t$
provides the evolution equation for the magnetic field
\begin{equation}
\label{evB}
\partial_t B^i + {\cal L}_{\mathbf{N}} B^i =
- (D \times N E)^i + NK B^i \; .
\end{equation}

A self-consistency check of Eq. \eqref{evE} can be performed by noting
that Eq. \eqref{Maxcov1} implies the charge conservation equation
\begin{equation}
\label{CQ}
\nabla_c j^c=0 \; .
\end{equation}
This equation can be written in terms of the 3+1 language as follows
(see Appendix):
\begin{equation}
\label{chargecons}
\partial_t \rho + {\cal L}_{\mathbf{N}} \rho =
- D_a (N \,^{(3)} j^a ) + N \rho K \; .
\end{equation}
Therefore, by replacing in the above equation the values of $\rho$ and
$\,^{(3)} j^c$ given by the Eqs. \eqref{Max3} and \eqref{evE}
respectively, one finds (after some algebra that involves the use of
the commutator of covariant derivatives applied to a vector field as
well as the Gauss-Codazzi equations) an identity.  In a similar way,
one can also check the self-consistency of Eqs. \eqref{Max4} and
\eqref{evB}.

At this point it is important to mention that a set of 3+1 Maxwell
equations analogous to Eqs.~\eqref{Max3}, \eqref{Max4}, \eqref{evE}
and \eqref{evB}, as well as the 3+1 charge conservation
Eq. \eqref{chargecons}, were derived previously by Thorne \& Macdonald
in~\cite{Thorne82} using the same sign conventions but a different
notation~\footnote{In order to compare equations (3.4a)$-$(3.4d) of
  Thorne \& Macdonald \cite{Thorne82} with ours we must translate
  their notation to ours as follows: $U_\alpha\rightarrow n_a$,
  $\theta\rightarrow -K$, \mbox{$\sigma_{\alpha\beta}\rightarrow
    -\left( K_{ab} -\frac{1}{3} K h_{ab}\right)$} which corresponds to
  minus the traceless part of the extrinsic curvature; $\tilde{\nabla}
  \rightarrow D$, $\alpha \rightarrow N$, $D_\tau M^\beta \rightarrow
  n^a \nabla_a M^b - M_a a^a n^b$ [cf. their Eq. (2.12)] where in the
  present case $M^a$ stands for $E^a$ or $B^a$. When multiplied by $N$
  the terms $n^a \nabla_a M^b - M_a a^a n^b$ can be easily arranged to
  give the first term on the l.h.s of Eq. \eqref{DynE} or the first two terms on
  the l.h.s of Eqs. \eqref{evE} and \eqref{evB}.  In general a tilde $\tilde{}$
  used by those authors under several quantities or symbols stands for
  quantities defined on $\Sigma_t$, whether they are 3-vectors or
  derivatives compatible with the 3-metric.}. These authors in turn
used the 3+1 congruence formalism of Maxwell's equations derived by
Ellis~\cite{Ellis73}.

It is important to note that in a flat spacetime Eqs.~\eqref{Max3},
\eqref{Max4}, \eqref{evE} and \eqref{evB} reduce to the familiar form
of Maxwell's equations. For the present case of a curved spacetime,
one can in fact rewrite equation~\eqref{Max3} in integral form as
\begin{equation}
\int_{\partial \Sigma_t} E^a \sqrt{h} \: \sigma_a d\sigma =
\int_{\Sigma_t} - n_a j^a \sqrt{h} \: dx^1 dx^2 dx^3 \; .
\end{equation}
On the left hand side (l.h.s), one has the flux of the electric field
lines across a closed two-surface lying on $\Sigma_t$ with normal $\sigma_a
\in T^{\Sigma_t}_p$.  On the right hand side (r.h.s), one has the
total charge measured by the Eulerian observers contained in the
volume enclosed by the two-surface.  The r.h.s is a consequence of
Eq. \eqref{CQ}, which implies that the total electric charge $Q$ is
conserved:
\begin{eqnarray}
Q &=& \int_{\Sigma_t} N j^t \sqrt{h} \: dx^1 dx^2 dx^3 \nonumber \\
&=& \int_{\Sigma_t} - n_a j^a \sqrt{h} \: dx^1 dx^2 dx^3 \nonumber \\
&=& \int_{\Sigma_t} \rho \sqrt{h} \: dx^1 dx^2 dx^3 \; . 
\end{eqnarray}
Note that in all the above expressions there appears the proper volume
element $\sqrt{h} \: dx^1 dx^2 dx^3$ on $\Sigma_t$ as measured by the
Eulerian observers.  A similar analysis can be done on
Eq.~\eqref{Max4}, except that in this case there are no magnetic
charges (the magnetic field lines are always closed).

As concerns the evolution equations \eqref{evE} and \eqref{evB}, some
of the extra terms appearing there are due to curvature effects plus
the fact that all observables are referred to the Eulerian
observers. For instance, if one uses during the evolution the
so-called {\em maximal slicing condition}\/ (which is defined by
imposing $K \equiv 0 \equiv \partial_t K$; this leads to an elliptic
equation for the lapse $N$ [cf. Eq.~\eqref{EDEfnew}]), then the terms
proportional to $N K$ on the r.h.s of Eqs. \eqref{evE} and \eqref{evB}
vanish identically. Otherwise, those terms are present and are
associated with the time variation of the proper volume elements on
$\Sigma_t$. Now, quite independently of the choice of a particular
time slicing, Thorne \& Macdonald have provided geometrical
interpretations of the extra terms that couple gravity with
electromagnetism in a non-trivial fashion. Such interpretations can
become even clearer when writing the evolution equations in integral
form~\cite{Thorne82}.

%%%%%%%%%%%%%%%%%%%%%%%%%
%%%   HYPERBOLICITY   %%%
%%%%%%%%%%%%%%%%%%%%%%%%%

\subsection{Hyperbolicity analysis of Maxwell's equations in curved
spacetimes}
\label{sec:hyperbolicity}

The system of evolution equations~\eqref{evE} and~\eqref{evB} for the
electric and magnetic fields can clearly be written as~\footnote{In
  Fourier space the system reads $\omega {\hat {\vec u}} -
  \mathbb{M}^i k_i {\hat {\vec u}}= {\hat {\vec {\cal S}}}.$}
\begin{equation}
\partial_t {\vec u} + \mathbb{M}^i \partial_i {\vec u} = {\vec S}\,\,\,.
\end{equation}
where ${\vec u}= (E^i,B^i)$ are the fundamental variables and
$\mathbb{M}^i$ are the characteristic matrices along the directions
$x^i$.

In order to gain some insight on the hyperbolic structure of Maxwell's
equations, let us first focus on the case of a flat spacetime
background in Cartesian coordinates and in vacuum ({\em i.e.} in the
absence of electric charges and currents). In such a case 
the matrix $\mathbb{M}^i$ is a $6\times 6$ block antidiagonal matrix which 
can be written as the following direct sum of two $3\times 3$ matrices:
\be
 \mathbb{M}^{i} = \mathbb{M}^{i}_{\rm up} \tilde{\oplus} \mathbb{M}^{i}_{\rm low}  \,\,\,,\\
\ee
where the symbol $\tilde{\oplus}$ means a direct sum by placing the blocks in the antidiagonal:
\begin{equation}
\label{Matcar}
\mathbb{M}^i \,\,=\,\, \left(
\begin{array}{cc}
         0 & \mathbb{M}^{i}_{\rm up}  \\
         \mathbb{M}^{i}_{\rm low} & 0        
\end{array}\right) \,\,\,.
\end{equation}
The components of the upper and lower matrices $\mathbb{M}^{i}_{\rm up}$ and $\mathbb{M}^{i}_{\rm low}$ are given 
respectively by 
\bea
 \mathbb{M}^{ilm}_{\rm up} &=& \epsilon^{ilm}_F \,\,\,\,(1\le l,m \le 3) \,\,\,,\\
\label{Mlow}
 \mathbb{M}^{ilm}_{\rm low} &=& -\epsilon^{ilm}_F \,\,\,\,(1\le l,m \le 3) \,\,\,.
\eea
The minus ``$-$'' in Eq. (\ref{Mlow}) corresponds to the asymmetry in sign in the dynamic Maxwell equations
(\ref{evE}) and (\ref{evB}). However, since $ \mathbb{M}^{ilm}_{\rm up} =  \mathbb{M}^{iml}_{\rm low}$ 
this shows that $\mathbb{M}^i$ is indeed symmetric [{\em e.g.} see Eq. (\ref{MatcarX}) ]. Therefore one concludes that
the system of equations (\ref{evE}) and (\ref{evB}) are in fact {\it  symmetric hyperbolic} and so they admit a well 
posed Cauchy problem.

One can further analyze the different modes of the characteristic
matrix $\mathbb{M}^i$. Let us take first the simple case where $E^i$
and $B^i$ consists of plane waves moving in the `x' direction so that
\begin{eqnarray}
E^i &=& {\hat E}^i e^{\imath(\omega t - kx)} \; , \\
B^i &=& {\hat B}^i e^{\imath(\omega t - kx)} \; .
\end{eqnarray}
Equations \eqref{evE} and \eqref{evB} then reduce to the following
algebraic system (remember that we are considering flat spacetime in
vacuum)
\begin{eqnarray}
\omega {\hat E}^i + \epsilon^{ijl}k_j {\hat B}^l &=& 0 \; ,\\
\omega {\hat B}^i - \epsilon^{ijl}k_j {\hat E}^l &=& 0 \; .
\end{eqnarray}
The matrix $\mathbb{M}^x $ then takes the form
\begin{equation}
\label{MatcarX}
\mathbb{M}^x \,\,=\,\, \left(
\begin{array}{cccccc}
         0 & 0 & 0 & 0 & 0 & 0  \\
         0 & 0 & 0 & 0 & 0 & c  \\
         0 & 0 & 0 & 0 &-c & 0  \\
         0 & 0 & 0 & 0 & 0 & 0  \\
         0 & 0 &-c & 0 & 0 & 0  \\
         0 & c & 0 & 0 & 0 & 0   
\end{array}\right) \,\,\,,
\end{equation} 
where $c:= k/\omega$ corresponds to the speed of light. The
constraints equations~\eqref{Max3} and~\eqref{Max4} impose the
following conditions
\begin{equation}
\label{constr_pw}
 {\hat E}_x =0 =  {\hat B}_x \; .
\end{equation}

Now, the eigenvalues of $\mathbb{M}^x $ are $\vec\lambda= (0,0,\pm
c,\pm c)$.  The eigenvalue $\lambda=0$ corresponds to the eigenvectors
${\vec e}_1 = (1,0,0,0,0,0)$ and ${\vec e}_2 = (0,0,0,1,0,0)$, the
eigenvalue $\lambda=-c$ corresponds to the eigenvectors \mbox{${\vec
    e}_3 = (0,-1,0,0,0,1)$} and ${\vec e}_4 = (0,0,1,0,1,0)$, and the
eigenvalue $\lambda=c$ corresponds to the eigenvectors \mbox{${\vec
    e}_5 = (0,1,0,0,0,1)$} and ${\vec e}_6 = (0,0,-1,0,1,0)$.

The determinant of the eigenvector matrix $\mathbb{R}^x$ is
\begin{equation}
\label{hypc}
{\rm det} \: (\mathbb{R}^x) = -4 \; .
\end{equation}
The set of eigenvectors therefore is complete, and the evolution
system turns to be strongly hyperbolic. This of course is not a surprise
since we already knew that $\mathbb{M}^x $ is symmetric and therefore
the system is in fact symmetric hyperbolic.

The first two modes with speed zero ($\lambda=0$) are clearly
unphysical since they are associated with modes that violate the
constraints~\eqref{constr_pw}.  On the other hand, the modes with
speed $\pm c$ do satisfy the constraints. These physical modes
correspond to ${\vec e}_3,{\vec e}_4, {\vec e}_5,{\vec e}_6$, and are
associated with the two polarizations states (each one with speed $\pm
c$) of the electromagnetic waves.

\bigskip

Now, for the case of waves propagating along an arbitrary direction
defined by the unit vector $\vec{s}$, the principal symbol of the
system is given by $\mathbb{C}:= \mathbb{M}^i s_i $.  Then we have
\begin{equation}
\mathbb{C} \,\,=\,\, \left(
\begin{array}{cccccc}
         0 & 0 & 0 & 0 & s_3 & -s_2  \\
         0 & 0 & 0 & -s_3 & 0 & s_1  \\
         0 & 0 & 0 & s_2 &-s_1 & 0  \\
         0 & -s_3 & s_2 & 0 & 0 & 0  \\
         s_3 & 0 &-s_1 & 0 & 0 & 0  \\
         -s_2 & s_1 & 0 & 0 & 0 & 0   
\end{array}\right) \,\,\,.
\end{equation} 
The eigenvalues are $\vec\lambda= (0,0,\pm 1,\pm 1)$ with
corresponding eigenvectors (from now on we will take the speed
of light to be equal to 1):
\begin{itemize}
\item $\lambda=0$:
\begin{eqnarray}
\label{evec1}
{\vec e}_1 &=& (s_1/s_3,s_2/s_3,1,0,0,0) \; , \\
{\vec e}_2 &=& (0,0,0,s_1/s_3,s_2/s_3,1) \; .
\end{eqnarray}
\item $\lambda=-1$
\begin{eqnarray}
{\vec e}_3 &=& (s_2,-(s_1^2+s_3^2)/s_1,s_2 s_3/s_1,-s_3/s_1,0,1) , \\
{\vec e}_4 &=& (-s_3,-s_2 s_3/s_1,(s_1^2+s_2^2)/s_1,-s_2/s_1,1,0) .
\hspace{10mm}
\end{eqnarray}
\item $\lambda=1$
\begin{eqnarray}
{\vec e}_5 &=& (-s_2,(s_1^2+s_3^2)/s_1,-s_2 s_3/s_1,-s_3/s_1,0,1) . 
\hspace{10mm} \\
\label{evec6}
{\vec e}_6 &=& (s_3,s_2 s_3/s_1,-(s_1^2+s_2^2)/s_1,-s_2/s_1,1,0) .
\end{eqnarray}
\end{itemize}
and the determinant of the eigenvector matrix $\mathbb{R}$ is
\begin{equation}
\label{hypc2}
{\rm det}\ \: (\mathbb{R}) = - 4 / (s_1^2 s_3^2) \; .
\end{equation}
The set of eigenvectors is clearly complete, and the evolution system
again turns out to be strongly hyperbolic. Since $\mathbb{C} $ is
symmetric, the system is in fact symmetric hyperbolic.

Again, the two modes with zero speed violate the constraints since
$\vec{S}\cdot {\vec e}_1$ and $\vec{S}\cdot {\vec e}_2$ do not vanish,
where $\vec{S}:=(s_1,s_2,s_3,s_1,s_2,s_3)$ (the constraints correspond
to $\partial_i u^i=0$, that is, ${\hat u}^i S_i= 0= {\hat E}^i s_i +
{\hat B}^i s_i$). On the other hand, one can see that $\vec{S}\cdot
{\vec e}_i$ vanish identically for $i=(3,...,6)$. Therefore such modes
satisfy the constraints and propagate with the speed of light. They
correspond to the two polarization states with speed $\pm 1$.

\bigskip

We can now consider the full system of evolution equations~\eqref{evE}
and~\eqref{evB} with a prescribed shift. The principal symbol now
reads,
\begin{equation}
\mathbb{C} = \left(
\begin{array}{cccccc}
         N_s & 0 & 0 & 0 &  {\cal N} s_3  & -  {\cal N} s_2 \\
         0 &N_s & 0 & -  {\cal N} s_3 & 0 &  {\cal N} s_1  \\
         0 & 0 & N_s &  {\cal N} s_2 &-  {\cal N} s_1 & 0  \\
         0 & -  {\cal N} s_3 &  {\cal N} s_2 & N_s & 0 & 0  \\
          {\cal N} s_3 & 0 &-  {\cal N} s_1 & 0 & N_s & 0  \\
         -  {\cal N} s_2 &  {\cal N} s_1 & 0 & 0 & 0 & N_s   
\end{array}\right) \; .
\end{equation} 
where $N_s:= N^i s_i$, and with ${\cal N}:= N/\sqrt{h}$ the {\em
  densitized}\/ lapse.

The eigenvalues of the principal symbol are now \mbox{$\vec{\lambda}=
  (N_s,N_s,N_s\pm {\cal N})$}. These correspond to the characteristic
speeds as measured by the Eulerian observers. In the flat spacetime
limit they reduce to those previously obtained. Their corresponding
eigenvectors are now
\begin{itemize}
\item $\lambda=N_s$
\begin{eqnarray}
\label{evec2}
{\vec e}_1 &=& (s_1/s_3,s_2/s_3,1,0,0,0) \; , \\
{\vec e}_2 &=& (0,0,0,s_1/s_3,s_2/s_3,1) \; .
\end{eqnarray}
\item $\lambda=N_s-{\cal N}$
\begin{eqnarray}
{\vec e}_3 &=& (s_2,-(s_1^2+s_3^2)/s_1,s_2 s_3/s_1,-s_3/s_1,0,1) . \\
{\vec e}_4 &=& (-s_3,-s_2 s_3/s_1,(s_1^2+s_2^2)/s_1,-s_2/s_1,1,0) .
\hspace{10mm}
\end{eqnarray}
\item $\lambda=N_s+{\cal N}$
\begin{eqnarray}
{\vec e}_5 &=& (-s_2,(s_1^2+s_3^2)/s_1,-s_2 s_3/s_1,-s_3/s_1,0,1) ,
\hspace{10mm} \\
\label{evec6c}
{\vec e}_6 &=& (s_3,s_2 s_3/s_1,-(s_1^2+s_2^2)/s_1,-s_2/s_1,1,0) .
\end{eqnarray}
\end{itemize}

Just as in the flat case, the first two modes which propagate with the
coordinate speed $N_s$ are unphysical since they violate the
constraints. On the other hand, the remaining modes satisfy the
constraints and propagate at the speed of light $N_s\pm {\cal N}$.

The characteristic speeds along a given direction $x^i$ can also be
written as \mbox{$\vec{\lambda}^i= (N^i,N^i,N^i\pm {\cal N}s^i )$}, so
that the projection along the $\vec{s}$ direction provides
$\vec{\lambda}$. This shows that physical modes propagate along the
light cones, corresponding to the coordinate speeds $N^i\pm {\cal
  N}s^i$.

%%%%%%%%%%%%%%%%%%%%%%%%%%%%%%%%%%
%%%   ENERGY-MOMENTUM TENSOR   %%%
%%%%%%%%%%%%%%%%%%%%%%%%%%%%%%%%%%

\subsection{The energy-momentum tensor of the electromagnetic field}
\label{sec:energymomentum}

The energy-momentum tensor of the EM field is given by
\begin{equation}
\label{TEM}
T_{ab} = \frac{1}{4\pi}\left[F_{ac}F^{\,\,\,c}_b 
- \frac{1}{4}g_{ab} \: F_{cd}F^{cd}\right] \; .
\end{equation}
Using Eq. (\ref{Fdcomp2}) we obtain
\begin{eqnarray}
F_{ac} F^{\,\,\,c}_b &=& -(E_a E_b + B_a B_b) \nonumber \\
&+& B^2 h_{ab} + E^2 n_a n_b
+ 2E^c B^d \,^{(3)} \,\!\epsilon_{cd(a}n_{b)} \; , \hspace{10mm}
\end{eqnarray}
where $E^2= E^a E_a$ and $B^2= B^a B_a$. 

From the above equation one can easily find
\begin{equation}
F_{ac} F^{ac}= -2 \left( E^2 -B^2 \right) \; .
\end{equation}
The energy-momentum tensor~\eqref{TEM} then becomes
\begin{eqnarray}
T_{ab} &=& \frac{1}{4\pi}\left[-(E_a E_b + B_a B_b)
+ \frac{1}{2} h_{ab} (E^2 +B^2) \right. \nonumber \\
&+& \left. \frac{1}{2}n_a n_b (E^2 + B^2)
+ 2 E^c B^d \,\,^{(3)} \,\!\epsilon_{cd(a}n_{b)} \right] \; ,
\hspace{10mm}
\end{eqnarray}
where we have used the fact that $g_{ab}= h_{ab} - n_a n_b$
[cf. Eq.~\eqref{proj}].  The 3+1 decomposition of this tensor is
[cf. Eq.~\eqref{Hab1}]
\begin{equation}
\label{TEM3+1}
T_{ab}= {\cal E} n_a n_b + n_a J_b + J_a n_b + S_{ab} \; ,
\end{equation}
where now [cf. Eqs. \eqref{3H}$-$\eqref{Hab5}]
\begin{eqnarray}
\label{Edens}
{\cal E} &:=& n^a n^b T_{ab} = \frac{1}{8\pi} (E^2 +B^2) \; , \\
\label{Poynt}
J_a &:=& - h_a^{\,\,\,c} n^d  T_{dc} = \frac{1}{4\pi} \,\,^{(3)}
\,\!\epsilon_{acd} E^c B^d \; , \\
\label{Stressten}
S_{ab} &:=& h_a^{\,\,\,c} h_b^{\,\,\,d} T_{cd} \nonumber \\
&=& \frac{1}{8\pi} \left[h_{ab}(E^2 +B^2)
- 2 (E_a E_b + B_a B_b) \right] . \hspace{8mm}
\end{eqnarray}
We identify $ {\cal E}$ with the energy-density of the EM field as
measured by the Eulerian observers, $J_a$ with the momentum density
measured by those observers (the {\em Poynting vector}), and $S_{ab}$
with the stress tensor~\footnote{Modulo notation,
  Eqs. \eqref{Edens}$-$\eqref{Stressten} coincide exactly with Eqs
  (3.10) of Thorne \& Macdonald \cite{Thorne82}.}.

Since the trace $T^a_{\,\,\,a}$ of the energy-momentum
tensor~\eqref{TEM} vanishes, then Eq.~\eqref{TEM3+1} leads to
[cf. Eqs. \eqref{Edens} and \eqref{Stressten}],
\begin{equation}
\label{nullT}
S  = {\cal E} \; ,
\end{equation}
where $S= S^a_{\,\,\,a}$ is the trace of the 3-tensor
(\ref{Stressten}).

%%%%%%%%%%%%%%%%%%%%%%%%%%%%
%%%   EINSTEIN-MAXWELL   %%%
%%%%%%%%%%%%%%%%%%%%%%%%%%%%

\subsection{The Einstein-Maxwell system}
\label{sec:EinsteinMaxwell}

We now consider the Einstein equations in 3+1 form, with the matter
sources provided by the electromagnetic field contributions of
Sec.~\ref{sec:energymomentum}.

The Hamiltonian and momentum constraints are,
respectively~\cite{York79,Gourgoulhon07,Alcubierre08a}
\begin{equation}
\label{CEHf}
^3 R + K^2 -  K_{ij}  K^{ij}= 16 \pi {\cal E} \; ,
\end{equation}
\begin{equation}
\label{CEMf}
D_l K^{il} - D^i K = 8 \pi J^i \; .
\end{equation}
The dynamic Einstein equations read~\cite{York79,Gourgoulhon07,Alcubierre08a}
\begin{eqnarray}
&& \partial_t K_{ij} + {\cal L}_{\mathbf{N}} K_{ij} + D_i D_j N
\nonumber \\
&& \hspace{5mm} - N \left( \,^3 R_{ij} + K K_{ij}  - 2 K_{il} K_j^l \right)
\nonumber \\
&& \hspace{5mm} = 4 \pi N \left[ h_{ij} (S-{\cal E}) - 2 S_{ij} \right]
= - 8\pi N S_{ij} \; , \hspace{10mm}
\label{EDEfnew}
\end{eqnarray}
where we used Eq.~\eqref{nullT} to simplify the r.h.s of
Eq.~\eqref{EDEfnew}.  Taking now the trace in Eq.~\eqref{EDEfnew}, and
using Eq.~\eqref{CEHf}, one obtains the following evolution equation
which can be very useful in many cases ({\em e.g.} see
Sec.~\ref{sec:initial} below):
\begin{eqnarray}  
&& \partial_t K + N^l \partial_l K + D^2 N -N K_{ij} K^{ij} \nonumber \\
&& \hspace{10mm} = 4 \pi N \left( S + {\cal E} \right) = 8 \pi N {\cal E} \; ,
\label{EDK}
\end{eqnarray}
where we used Eq.~\eqref{nullT} in the last step. Here $D^2$ stands
for the Laplacian operator compatible with the 3-metric $h_{ab}$.

In all the four equations~\eqref{CEHf}-\eqref{EDK}, the r.h.s is
given in terms of the energy-momentum contributions defined in
Eqs.~\eqref{Edens}-\eqref{Stressten}.

It is well known that the evolution equations~\eqref{EDEfnew}, when
written in first order form, are only weakly hyperbolic (see
Ref.~\cite{Alcubierre08a} for a thorough review), and so do not admit
a well-posed Cauchy problem (in the Hadamard sense). However, by
adding suitable multiples of the constraints and using a conformal
decomposition ({\em e.g.} the BSSN
formulation~\cite{Baumgarte:1998te}) one can write the evolution
system in such a way that the evolution system admits a well posed
Cauchy problem.  Here we are not concerned with with that issue, but
will will consider it when we study numerical evolutions of the
Einstein-Maxwell system in a future paper.

The Cauchy problem in this case can be summarized as follows: given
the initial data $(\Sigma_t, h_{ab},K_{ab}, E_a, B_a)$, satisfying the
constraints~\eqref{CEHf}, \eqref{CEMf}, \eqref{Max3} and \eqref{Max4},
one can then evolve forward in time (given a prescription for $N$ and
$N^i$) the fields $h_{ab},K_{ab}, E_a, B_a$ using their evolution
equations~\eqref{K_ab}, \eqref{EDEfnew} (or their corresponding
strongly hyperbolic formulated equations), \eqref{evE} and \eqref{evB}
respectively~\footnote{See Refs. \cite{Lichnerowicz1955,Witten1960,Wald84},
  for a review and discussion on the initial value problem of the
  Einstein-Maxwell system.}.

%%%%%%%%%%%%%%%%%%%%%%
%%%   POTENTIALS   %%%
%%%%%%%%%%%%%%%%%%%%%%

\subsection{Electromagnetic potentials}
\label{sec:potentials}

Up until this point we have worked directly with the electric and
magnetic fields and ignored the potentials.  This is quite deliberate,
as the use of the potentials can complicate matters, in particular due
to the fact that one needs to choose a gauge. Also, the evolution
equations for the electromagnetic field when written in terms of the
potentials are second order in space and time, which brings extra
complications coming from the fact that covariant derivatives do not
commute. Nevertheless, here we will very briefly describe, without
going into any details, the 3+1 form of the electromagnetic potentials
and their relation with the electric and magnetic fields.

Let us start by remembering that in terms of the potential 4-vector
$A^a$ the Faraday tensor is given by equation~\eqref{Faraday}, which
we rewrite here for concreteness:
\begin{eqnarray}
F_{ab} = \partial_a A_b - \partial_b A_a \; .
\end{eqnarray}
Starting from the potential 4-vector, one can now define a 3+1
``scalar potential'' $\Phi$ through
\begin{equation}
\Phi := - n_a A^a \; ,
\end{equation}
together with a potential 3-vector $^{(3)} \!A^a$ defined as
\begin{equation}
^{(3)} \!A^a := h^a_{\,\,b} A^b \; .
\end{equation}
From these definitions one can inmediately find that, in a
coordinate system adapted to the 3+1 foliation:
\begin{equation}
\Phi = N A^t = - \frac{1}{N} \: \left( A_t + N^a A_a \right) \; ,
\end{equation}
and
\begin{equation}
^{(3)} \!A^a = A^a - n^a \Phi \; , \qquad
^{(3)} \!A_i = A_i \; .
\end{equation}

Now, by projecting the expression for the Faraday tensor in terms of
$A^a$ given above one can obtain, after some algebra, the following
relation between the electric-field $E_i$ and the 3+1 potentials:
\begin{equation}
\partial_t \,^{(3)} \!A_i + {\cal L}_{\mathbf{N}} \,^{(3)} \!A_i
= - N E_i - D_i \left( N \Phi \right) \; .
\label{eq:dta}
\end{equation}
Notice that this equation can in fact be interpreted as an evolution
equation for the potential 3-vector $\,^{(3)} \!A_i$.

Similarly, one can also obtain the following expression for the
magnetic field $B_i$ in terms of $\,^{(3)} \!A_i$:
\begin{eqnarray}
B^i &=& \frac{1}{2} \, ^{(3)}\,\!\epsilon^{imn}
\left( \partial_m \,^{(3)} \!A_n - \partial_n \,^{(3)} \!A_m \right) \nonumber \\
&=& \, ^{(3)}\,\!\epsilon^{imn} \partial_m \,^{(3)} \!A_n
= \left( D \times \,^{(3)} \!A \right)^i \; ,
\label{eq:rota}
\end{eqnarray}
with the rotational operator $(D \times \,^{(3)} \!A)$ defined in the same way as
before.

At this point, one could take the point of view that the independent
dynamical variables are in fact $\,^{(3)} \!A^i= h^{ij}\,^{(3)} \!A_j$ 
and $E^i$, with evolution
equations given by~\eqref{evE} and~\eqref{eq:dta}, and simply define
the magnetic field through Eq.~\eqref{eq:rota}, therefore ignoring
Eq.~\eqref{evB} (which is now just a consequence of the definition of
$B^i$).  One would also find that the magnetic constraint~\eqref{Max4}
is now trivial.  We would then have a system of evolution equations
that is first order in time and second order in space, with only one
constraint, namely the electric constraint~\eqref{Max3}.

Of course, one would still have to choose a gauge condition in order
to evolve the scalar potential $\Phi$. One possibility would be to
take the {\em Lorentz gauge}, which is given in terms of the 4-vector
potential as:
\begin{equation}
\nabla_a A^a = 0 \; .
\label{eq:Lorentz1}
\end{equation}
This gauge condition can be easily seen to take the following form in
3+1 language
\begin{equation}
\partial_t \Phi + {\cal L}_{\mathbf{N}} \Phi =
- D_m \left( N \,^{(3)} \!A^m \right) + N K \Phi \; ,
\label{eq:Lorentz2}
\end{equation}
with $K$ the trace of the extrinsic curvature, which clearly provides
us with an evolution equation for $\Phi$.

Taking this point of view, however, has one serious drawback.  One can
show that the system of evolution equations given by~\eqref{evE},
\eqref{eq:dta} and \eqref{eq:Lorentz2} is in fact only weakly
hyperbolic even in flat space ({\em i.e.} it has real eigenvalues, but
does not have a complete set of eigenvectors), so that the system is
not well-posed.  There are certainly ways around this, involving
defining new auxiliary variables and {\em crucially}\/ commuting the
second covariant derivatives of $\,^{(3)} \!A^i$ that would appear in Eq.~\eqref{evE}
when we write the magnetic field as~\eqref{eq:rota} (which
considerably complicates the equation by bringing in a contribution
from the Riemann tensor), but we will not go into such details here
(see {\em e.g.} Ref.~\cite{Knapp:2002fm}).  It is enough to say that,
even though strongly hyperbolic versions of the evolution system for
$\,^{(3)} \!A^i$ and $E^i$ do exist, it is simpler and much cleaner to work with
the electric and magnetic fields directly, and consider the potentials
just as auxiliary variables when (and if) they are needed.

%%%%%%%%%%%%%%%%%%%%%%%%%%%%%%%%%%%%%%%%%%%%%%%%%%%%%%%%%
%%%   INITIAL DATA FOR MULTIPLE CHARGED BLACK HOLES   %%%
%%%%%%%%%%%%%%%%%%%%%%%%%%%%%%%%%%%%%%%%%%%%%%%%%%%%%%%%%

\section{Initial data for multiple charged black holes}
\label{sec:initial}

In this Section we will consider the problem of finding suitable
initial data for multiple charged black holes.  For simplicity, will
concentrate on the case of time-symmetric initial data for which the
extrinsic curvature vanishes $K_{ab}=0$. Furthermore, we will also
assume that the initial magnetic field is zero, in which case the
momentum constraints are identically satisfied.

The problem of solving the Einstein-Maxwell constraint equations was
studied previously by Bowen in~\cite{Bowen85}, where he used a method
of images (notably for the electric field) to construct a manifold
that represents two isometric asymptotically flat universes with $n$
throats connecting them. In that work Bowen describes a solution for
the electric field that is inversion symmetric through the throats
(but which do not arise from a potential), which then allows one to
find numerical initial data for the 3-metric that represents $n$
charged black holes. Here, however, we will use a different approach
and will look instead for solutions that are not inversion symmetric,
but that rather represent a series of throats that connect our
universe to $n$ {\em distinct}\/ asymptotically flat universes, more
in the spirit of the Brill-Lindquist~\cite{Brill63} initial data for
time-symmetric black holes, or the Brandt-Bruegmann~\cite{Brandt97}
puncture data for spinning or moving black holes.

%%%%%%%%%%%%%%%%%%%%%%%%%%%%%%%%%%%%
%%%   CONFORMAL TRANSFORMATION   %%%
%%%%%%%%%%%%%%%%%%%%%%%%%%%%%%%%%%%%

\subsection{Conformal transformation of the metric and electric field}
\label{sec:conformal}

For time symmetric initial data and vanishing magnetic field, the
problem of finding initial data reduces to finding a solution of
the electric constraint~\eqref{Max3}
\begin{equation}
D_a E^a = 4 \pi \rho \; ,
\label{eq:electricconstraint}
\end{equation}
together with the Hamiltonian constraint~\eqref{CEHf}
\begin{equation}
^3 R = 16 \pi {\cal E} \; ,
\end{equation}
with the energy density of the EM given by:
\begin{equation}
{\cal E} = \frac{1}{8\pi} \: E_a E^a \; .
\end{equation}

Let us now assume that the spatial metric $h_{ab}$ is conformally
flat, so that we can rewrite it as:
\begin{equation}
h_{ab} = \psi^4 \hat{h}_{ab} \; ,
\end{equation}
with $\psi$ the conformal factor and $\hat{h}_{ab}$ a flat background
metric in arbitrary coordinates. The Hamiltonian constraint then
reduces to the following elliptic equation for the conformal factor
\begin{equation}
\hat{D}^2 \psi + \frac{1}{4} \: \psi^5 E_a E^a = 0 \; ,
\end{equation}
with $\hat{D}^2$ the Laplacian operator compatible with the background
metric.

With the above conformal transformation in mind, let us now consider
the electric constraint~\eqref{eq:electricconstraint}.  Notice first
that $D_a$ is the derivative operator associated with the physical
metric $h_{ab}$.  Notice also that, quite generally, for any vector
$v^a$ we have
\begin{equation}
D_a v^a = \hat{D}_a v^a + 6 v^a \partial_a \ln \psi \; ,
\end{equation}
with $\psi$ the conformal factor introduced above and $\hat{D}$ the
derivative operator associated with the conformal
metric $\hat{h}_{ab}$.  This implies in particular that
\begin{equation}
D_a \left( \psi^n v^a \right) = \psi^n \left[
\hat{D}_a v^a + (6+n) v^a \partial_a \ln \psi \right] \; .
\end{equation}

Using this result it is then natural to define the conformally
rescaled electric field as
\begin{equation}
\label{Econf}
\hat{E}^a := \psi^6 E^a \; , \qquad \hat{E}_a := \psi^2 E_a \; ,
\end{equation}
and the conformally rescaled charge density as
\begin{equation}
\hat{\rho} := \psi^6 \rho \; .
\end{equation}
The electric constraint then reduces to
\begin{equation}
\hat{D}_a \hat{E}^a = 4 \pi \hat{\rho} \; ,
\label{eq:electricconf}
\end{equation}
where now the divergence is calculated with respect to the conformal
metric.

In terms of the conformal electric field just defined, the Hamiltonian
constraint takes the final form:
\begin{equation}
\hat{D}^2 \psi + \frac{1}{4 \psi^3} \: \hat{E}_a \hat{E}^a = 0 \; .
\label{eq:hamconf}
\end{equation}

In order to find initial data, one must then first solve the conformal
electric constraint~\eqref{eq:electricconf}, and then plug in the
solution for $\hat{E}^a$ into the Hamiltonian constraint in order to
solve for the conformal factor $\psi$.

In fact, Eq.~\eqref{eq:hamconf} can also be written as
\begin{equation}
{\bar \psi} \hat{D}^2 {\bar \psi}
- \frac{1}{2} (\hat{D}^a {\bar \psi}) (\hat{D}_a {\bar \psi})
+ \frac{1}{2} \: \hat{E}_a \hat{E}^a = 0 \; .
\label{eq:hamconf2}
\end{equation}
where ${\bar \psi}:= \psi^2$. This equation was considered by
Bowen~\cite{Bowen85} for solving the initial data for the single
charged black hole case (see the analysis below).

%%%%%%%%%%%%%%%%%%%%%%%%%%%%%%%%%
%%%   ANALYTIC INITIAL DATA   %%%
%%%%%%%%%%%%%%%%%%%%%%%%%%%%%%%%%

\subsection{Exact initial data  multiple charged black holes with the
same charge-to-mass ratio}
\label{sec:exactdata}

In order to find initial data for multiple charged black holes we will
first assume that the background metric is flat. Let us introduce a
conformal electric potential $\varphi$ such that
\begin{equation}
\label{hatE_a}
\hat{E}_a = - \partial_a \varphi \; .
\end{equation}
Do notice that the conformal potential $\varphi$ does not coincide with
the physical potential $\Phi$ discussed in Sec.~\ref{sec:potentials}
above, since even in the absence of a vector potential
equation~\eqref{eq:dta} clearly shows that the relation between $\Phi$
and the physical electric field involves the lapse function.

Using Eq.~\eqref{hatE_a}, the electric constraint can be rewritten as
\begin{equation}
\hat{D}^2 \varphi = - 4 \pi \hat{\rho} \; .
\label{eq:electricconf2}
\end{equation}
From now on we will also assume that we are in a region away from any
charges, so that $\hat{\rho}=0$.

Before attempting to find initial data for multiple charged black
holes, let us recall for a moment the Reissner-Nordstr\"om analytic
static solution for a single charged black hole with mass $M$ and
charge $Q$~\cite{Reissner16,Nordstrom18}, for which the conformal
electric potential $\varphi$ and the conformal factor $\psi$ are given
by:
\begin{equation}
\label{RNsols}
\varphi = \frac{Q}{r} \; , \qquad
\psi = \left[ \left( 1 + \frac{M}{2r} \right)^2
- \frac{Q^2}{4 r^2} \right]^{1/2} , 
\end{equation}
where we have assumed that the black hole is centered on the origin of
the coordinate system $r=0$.  

The conformal and physical electric fields for this solution are
purely radial and are given by:
\begin{equation}
\label{E_r}
\hat{E}^r = \frac{Q}{r^2} \; , \qquad
E^r = \frac{Q}{r^2 \psi^6} \; .
\end{equation}
The fact that the conformal factor $\psi$ above is an exact solution
of Eq.~(\ref{eq:hamconf}) for this electric field can be verified by
direct substitution.

Since in this case the spacetime is static, Eq.~\eqref{EDK} provides a
linear elliptic equation for the lapse:
\begin{equation}  
\label{EDKstat}
D^2 N = 8 \pi N {\cal E} = N E^a E_a  \; ,
\end{equation}
which in terms of the conformal variables reads
\begin{equation}  
\label{EDKstatconf}
\hat{D}^2 N + \frac{2}{\psi} (\hat{D}^i N) (\hat{D}_i \psi)
= \frac{N}{\psi^4} \: \hat{E}_a \hat{E}^a  \; .
\end{equation}
One can now also confirm by direct substitution that the lapse given
by Eq.~\eqref{LapseRN} below solves Eq.~\eqref{EDKstatconf} when
using in turn Eqs.~\eqref{RNsols} and \eqref{E_r}:
\begin{equation}
\label{LapseRN}
N = \frac{(1+M/2r)(1-M/2r) + Q^2/4r^2}{(1+M/2r)^2 - Q^2/4r^2} \; .
\end{equation}
Notice that Eqs.~\eqref{RNsols}, \eqref{E_r} and \eqref{LapseRN}
correspond to the Reissner-Nordstr\"om solution in isotropic ({\em
  i.e.}  conformally flat) coordinates, and not in the standard
Schwarzschild-like coordinates one finds in most text
books \footnote{In Schwarzschild (area) coordinates the
  Reissner-Nordstr\"om solution is given by $ds^2= -\left(1
  -\frac{2M}{{\bar r}}+ \frac{Q}{{\bar r}^2} \right) dt^2 + \left(1
  -\frac{2M}{{\bar r}}+ \frac{Q}{{\bar r}^2} \right)^{-1} d{\bar r}^2
  + {\bar r}^2 d\Omega^2$. }. It is clear that by taking $Q \equiv 0$,
the above solution reduces to the Schwarzschild solution in isotropic
coordinates.

\bigskip

Based on the form of the conformal factor for the Reissner-Nordstr\"om
solution given above, we will now propose the following {\em ansatz}\/
for the conformal factor in the presence of a generic electric
potential $\varphi$ that is a solution of the electric
constraint~\eqref{eq:electricconf2}:
\begin{equation}
{\bar \psi} = \psi^2 = \left( 1 + \eta \right)^2 - \frac{\varphi^2}{4}
\; .
\label{eq:ansatz1}
 \end{equation}
Substituting this back into equation~\eqref{eq:hamconf2} we find,
after some algebra, the following elliptic equation for $\eta$:
\begin{eqnarray}
\left( 1 + \eta \right) \left( \left( 1 + \eta \right)^2
- \frac{\varphi^2}{4} \right) \hat{D}^2 \eta \hspace{5mm} && \nonumber \\
- \frac{\varphi^4}{4} \: \partial_m \left( \frac{\eta}{\varphi} \right) 
\partial^m \left( \frac{\eta+2}{\varphi} \right) &=& 0 \; ,
\label{eq:hamconf3}
\end{eqnarray}
where we have already used the fact that $\hat{D}^2 \varphi = 0$.

We can now easily notice a remarkable fact: If we take the function
$\eta$ to be proportional to the electric potential $\varphi$, that is
$\eta = k \varphi$ for some constant $k$, then
equation~\eqref{eq:hamconf2} is identically satisfied since in such a
case we clearly have $\partial_m \left( \eta / \varphi \right)=0$ and
$\hat{D}^2 \eta=0$ (remember that $\varphi$ solves the electric
constraint away from charges).

We will now use this fact to find an exact solution of the Hamiltonian
constraint for multiple black holes. Let us assume that in the
conformal space we have a series of point charges with values $Q_i$
located at the points $\vec{r}_i$.  The solution for the potential
$\varphi$ is then clearly
\begin{equation}
\label{phimultsol}
\varphi = \sum_{i=1}^n \frac{Q_i}{|\vec{r} - \vec{r}_i |} \; .
\end{equation}
Let us now choose $\eta$ proportional to $\varphi$ in the following way:
\begin{equation}
\eta = k \varphi = k \sum_{i=1}^n \frac{Q_i}{|\vec{r} - \vec{r}_i |}
\equiv \sum_{i=1}^n \frac{M_i}{2 \: |\vec{r} - \vec{r}_i |} \; .
\end{equation}
We can now construct a conformal factor that satisfies the Hamiltonian
constraint as in~\eqref{eq:ansatz1}:
\begin{equation}
\psi^2 = \left( 1 + \sum_{i=1}^n \frac{M_i}{2|\vec{r}-\vec{r}_i |} \right)^2
- \frac{1}{4} \left( \sum_{i=1}^n \frac{Q_i}{|\vec{r}-\vec{r}_i |} \right)^2
\; ,
\label{eq:psiexact}
\end{equation}
with $M_i = 2 k Q_i$, and $k$ an arbitrary constant.  This solution
represents a series of $n$ charged black holes, all of which have {\em
  the same charge-to-mass ratio} $Q_i/M_i=1/2k$.

One could ask at this point how can we know that the above solution in
fact does represent a series of black holes.  There are several ways
to see that this should be so.  First, notice that if we take $n=1$
this is just the standard Reissner-Nordstr\"om solution for a single
charged black hole. Second, in the case when all the charges vanish
the conformal factor~\eqref{eq:psiexact} above reduces to the well
known Brill-Lindquist conformal factor for a series of non-charged
black holes~\cite{Brill63}. Also, if the different points $\vec{r}_i$
are very far apart, then close to each of them the conformal factor
again reduces essentially to the Reissner-Nordstr\"om solution, so we
would indeed have a series of $n$ charged black holes.  Finally,
notice that because of the singularities in the conformal factor, as
we approach each point $\vec{r}_i$ the areas of spheres centered
around that point first become smaller and then increase again.  That
is, our initial data is a topological construction with a series of
wormholes connecting to other asymptotically flat regions.  Since the
initial data is time symmetric the throats of these wormholes ({\em
  i.e.} the minimal surfaces) in fact correspond to apparent
horizons. Of course, if some of the points $\vec{r}_i$ are very close
to each other one could find common apparent horizons around them, so
that we actually have fewer black holes with complicated internal
topologies. As a final comment, notice also that due to the tidal
forces between the different black holes, the throats of the wormholes
can not be expected to be spherical, and their precise shape and
location should be found numerically.

That an exact solution of the Hamiltonian constraint for multiple
charged black holes exists at all is a surprising result.  Notice,
however, that we have only found a solution for the {\em initial
  data}. In general, one would expect this initial configuration to
evolve as each black hole reacts to the gravitational and electric
fields of the other black holes, so that a non-zero extrinsic
curvature and magnetic field would quickly develop.

There is in fact one notable exception to this.  In order to find it
we will first rewrite the conformal factor~\eqref{eq:psiexact} in the
following way
\begin{eqnarray}
\psi^2 &=& 1 + 2 k \sum_{i=1}^n \frac{Q_i}{|\vec{r} - \vec{r}_i |}
\nonumber \\
&+& \left( k^2 - \frac{1}{4} \right)
\left( \sum_{i=1}^n \frac{Q_i}{|\vec{r} - \vec{r}_i |} \right)^2 ,
\end{eqnarray}
where we have already used the fact that $M_i = 2 k \: Q_i$.  If we
now take $k=1/2$, which implies $M_i = Q_i$, then the conformal factor
reduces to:
\begin{equation}
\label{psisq}
\psi^2 = 1 + \sum_{i=1}^n \frac{Q_i}{|\vec{r} - \vec{r}_i |} \; ,
\end{equation}
which now corresponds to initial data for a series of {\em extremal}\/
black holes. Amazingly, one can show that this multi-extremal black
hole solution turns out to be static, that is, the gravitational
attraction of all the black holes is {\em exactly canceled out}\/ by
their electrostatic repulsion, so that the black holes never move from
their initial positions. The lapse function for such a static solution
is given by:
\begin{equation}
\label{Nmultstat}
N = 1/\psi^2 = \left[ 1 + \sum_{i=1}^n \frac{Q_i}{|\vec{r} - \vec{r}_i |}
\right]^{-1} .
\end{equation}
which solves Eq.~\eqref{EDKstatconf} exactly.

This multi-extremal static solution was first obtained by
Papapetrou~\cite{Papapetrou1945} and Majumdar~\cite{Majumdar1947} (see
\cite{Heusler1996} for a review), and was further analyzed by several
authors~\cite{Perjes1971,Israel1972,Hartle1972b,Chrusciel1994,Heusler1997}~\footnote{In
  order to recover the physical electric field for this static
  solution, one has to remember the definition \eqref{Electric}, so
  that $E_a = -n^b F_{ba}$. Since in this case the spacetime is static
  we have $n^a= (1/N)\left(\partial/\partial t\right)^a$, where
  $\left(\partial/\partial t\right)^a$ is the timelike Killing vector
  field which is hypersurface orthogonal. Then $E_a = -(1/N) F_{ta} =
  (1/N) \partial_a A_t$. Using Eq. \eqref{Nmultstat} one has $E_a =
  \psi^2 \partial_a A_t$. From Eqs. \eqref{Econf} and \eqref{hatE_a}
  one obtains $\partial_a A_t= -\psi^{-4} \partial_a \varphi$. Finally
  using $\psi^2 = 1 + \varphi$ [cf. Eqs. \eqref{phimultsol} and
    \eqref{psisq}] one concludes $A_t = const. + (1+ \varphi)^{-1} =
  const. + \psi^{-2}$, where the integration constant can be fixed in
  several ways. In the standard form of Papapetrou and Majumdar the
  constant is usually fixed to be zero. However one can choose it
  equal to -1 if one demands that $A_t$ vanish asymptotically.}. This
solution is so well known that it can even be found in some text books
(see {\em e.g.}\/ the recent book by Carroll~\cite{Carroll04a}). On
the other hand, as far as we are aware the exact solution of the
Hamiltonian constraint for non-extremal black holes with equal
charge-to-mass ratios presented above was not previously known.

%%%%%%%%%%%%%%%%%%%%%%%%%%%%%%%%%%
%%%   NUMERICAL INITIAL DATA   %%%
%%%%%%%%%%%%%%%%%%%%%%%%%%%%%%%%%%

\subsection{Numerical initial data for multiple charged black holes with
different charge-to-mass ratios}
\label{sec:numericaldata}

The exact multiple black hole solution of the Hamiltonian constraint
found in the previous section is only valid in the case when all the
black holes have the same charge-to-mass ratio.  When considering
different charge-to-mass ratios we have not been able to find a closed
form solution.  On the other hand, finding numerical solutions can be
done easily enough.  In order to do this we will first modify our
ansatz~\eqref{eq:ansatz1} above in the following way:
\begin{equation}
{\bar \psi} = \psi^2 = \left( u + \eta \right)^2 - \frac{\varphi^2}{4}
\; ,
\label{eq:ansatz2}
\end{equation}
with
\begin{eqnarray}
\eta &=& \sum_{i=1}^N \frac{M_i}{2|\vec{r} - \vec{r}_i |} \; , \\
\varphi &=& \sum_{i=1}^N \frac{Q_i}{|\vec{r} - \vec{r}_i |} \; ,
\end{eqnarray}
and where $u$ is a function that goes to 1 at infinity, and in fact is
identically equal to 1 everywhere in the case when all charge-to-mass
ratios are equal.

Substituting now Eq.~\eqref{eq:ansatz2} into the Hamiltonian
constraint we find the following elliptic equation for $u$:
\begin{eqnarray}
\left( \eta + u \right) \left( \left( \eta + u \right)^2
- \frac{\varphi^2}{4} \right) \hat{D}^2 u && \nonumber \\
&& \hspace{-43mm}- \frac{\varphi^4}{4} \:
\partial_m \left( \frac{\eta + u - 1}{\varphi} \right) 
\partial^m \left( \frac{\eta + u + 1}{\varphi} \right) = 0 \; ,
\hspace{10mm}
\label{eq:hamconf4}
\end{eqnarray}
The above equation needs to be solved numerically for $u$ in the case
when the charge-to-mass ratios of the different black holes are not
all equal.

Before presenting some examples of numerical solutions for the case of
two black holes, it is important to investigate the expected behaviour
of the function $u$ close to each of the ``punctures'', that is, close
to each of the points $\vec{r}=\vec{r}_i$.  In order to do this, let
us now use a system of spherical coordinates $(r,\theta,\phi)$ adapted
to one of the black holes.  Without loss of generality we will choose
that black hole as the one identified with the label 1, so that
$\vec{r}_1=0$.  Let us now examine the behavior of the different terms
in equation~\eqref{eq:hamconf4} for small $r$.  Consider first the
coefficient of the Laplacian operator:
\begin{equation}
T_1 := \left( \eta + u \right) \left( \left( \eta + u \right)^2
- \frac{\varphi^2}{4} \right)
 \; .
\end{equation}
Let us assume for the moment that $u$ is finite at each of the
punctures.  From the form of the functions $\eta$ and $\varphi$, it is
then clear that for small $r$ this term behaves in general as
\begin{equation}
T_1 \sim 1 / r^3 \; .
\end{equation}
Consider now the term with first order derivatives in Eq.~\eqref{eq:hamconf4}:
\begin{equation}
T_2 := \varphi^4 \: \partial_m \left( \frac{\eta + u -
  1}{\varphi} \right) \partial^m \left( \frac{\eta + u + 1}{\varphi}
\right) \; .
\end{equation}
In order to analyze the behaviour of this term for small $r$, we will
first expand the derivatives to obtain
\begin{eqnarray}
T_2 &=& \left[ \varphi \: \partial_m (\eta + u)
- (\eta+u-1) \: \partial_m \varphi \right] \nonumber \\
&& \left[ \varphi \: \partial^m (\eta+u)
- (\eta+u+1) \: \partial^m \varphi\right]
\; .
\end{eqnarray}
We will now rewrite the functions $\eta$ and $\varphi$ above as
\begin{equation}
\eta = \frac{M_1}{2r} + H(r,\theta,\phi) \; , \qquad
\varphi = \frac{Q_1}{r} + F(r,\theta,\phi) \; ,
\end{equation}
with $H$ and $F$ given by
\begin{equation}
H = \sum_{i \neq 1} \frac{M_i}{2|\vec{r} - \vec{r}_i |} \; , \qquad
F = \sum_{i \neq 1} \frac{Q_i}{|\vec{r} - \vec{r}_i |} \; ,
\end{equation}
which are clearly regular functions at $r=0$.  Substituting into $T_2$
and expanding we find, after some algebra, that for small $r$ this
term behaves as:
\begin{equation}
T_2 \sim 1/r^4 \; .
\end{equation}
Notice that naively one could expect $T_2$ to diverge as $1/r^6$, due
to the presence of terms of the form \mbox{$(\varphi \: \partial_r
  \eta)^2$}, but in fact all such terms cancel out and we are left
with a dominant divergence of order $1/r^4$.

From the behaviour of $T_1$ and $T_2$ for small $r$, we then find that
in order for equation~\eqref{eq:hamconf4} to be consistent the
Laplacian of $u$ must behave for small $r$ as:
\begin{equation}
\hat{D}^2 u \sim 1/r \; .
\end{equation}

Now, in spherical coordinates the (flat) Laplacian is given by
\begin{equation}
\hat{D}^2 u = \partial_r^2 u + \frac{2}{r} \: \partial_r u
+ \frac{1}{r^2} \: L^2 u \; ,
\end{equation}
with $L^2$ the angular operator
\begin{equation}
L^2 u := \frac{1}{\sin \theta} \: \partial_\theta \left(
\sin \theta \: \partial_\theta u \right)
+ \frac{1}{\sin^2 \theta} \: \partial_\varphi^2 u \; .
\end{equation}
This implies that in order to have $\hat{D}^2 u$ behaving as expected
for small $r$ we must ask for the function $u$ to have a Taylor
expansion near the origin of the form:
\begin{equation}
u = a + b(\theta,\phi) \: r \; ,
\end{equation}
with $a$ a constant and $b(\theta,\phi)$ some regular function of the
angular coordinates.  Notice that $b(\theta,\phi)$ {\em must}\/ be
non-zero for equation~\eqref{eq:hamconf4} to be consistent, so the
above expansion implies that $u$ is not regular at the origin
(remember that $r$ is a radial coordinate).  The function $u$ then
turns out to be only $C^0$ at the punctures, that is, it is finite and
continuous, but its derivatives are no longer continuous.  In other
words, the function $u$ is in general expected to have a kink ({\em
  i.e.} a conical singularity) at each of the punctures (we will see
in the numerical examples below that this is indeed the case).

%%%%%%%%%%%%%%%%%%%%%%%%%%%%%%
%%%   NUMERICAL EXAMPLES   %%%
%%%%%%%%%%%%%%%%%%%%%%%%%%%%%%

\subsection{Numerical examples for the case of two charged black holes}
\label{sec:examples}

We have constructed a simple numerical code to solve
equation~\eqref{eq:hamconf4} for the case of two charged black holes
with different charge-to-mass ratios.  In such a case one can locate
both black holes along the $z$ axis. The situation is then clearly
axisymmetric, so the problem is effectively two-dimensional.

As a boundary condition we ask for the function $u$ to behave as
\mbox{$u = 1 + c/r$} for large $r$, where $r=\sqrt{\rho^2 + z^2}$ and
$c$ is some constant.  In order to eliminate the unknown constant this
boundary condition is differentiated and applied in the following way:
\begin{equation}
\partial_r u = \frac{1-u}{r} \; .
\end{equation}

Our code uses cylindrical coordinates $(\rho,z,\phi)$ instead of
spherical coordinates $(r,\theta,\phi)$, so that in practice we assume
that far away the dependence on the azimuthal angle $\theta$ can be
ignored, so that one can write
\begin{equation}
\partial_\rho u = \left( \frac{\rho}{r} \right) \partial_r u \; , \qquad
\partial_z u = \left( \frac{z}{r} \right) \partial_r u \; .
\end{equation}
The final boundary condition on the $\rho$ boundaries is then
\begin{equation}
\left( \frac{r}{\rho} \right) \partial_\rho u = \frac{1-u}{r} \; ,
\end{equation}
with an analogous condition on the $z$ boundaries.

Since here we are mainly interested in showing that the solutions for
$u$ can be easily found and that they behave as expected, we have
decided to write a very simple code that instead of solving the
elliptic equation directly solves an associated hyperbolic problem
of the form:
\begin{eqnarray}
\partial^2_t u &=& \hat{D}^2 u - \frac{\varphi^4}{4} \: \partial_m
\left( \frac{\eta + u - 1}{\varphi} \right) \partial^m \left(
\frac{\eta + u + 1}{\varphi} \right) \nonumber \\
&& \times \left[ \left( \eta + u \right)
\left( \left( \eta + u \right)^2 - \frac{\varphi^2}{4} \right) \right]^{-1} ,
\label{eq:wavelike}
\end{eqnarray}
with $t$ a fictitious time parameter.  The boundary conditions are
then also modified to outgoing-wave boundary conditions of the form:
\begin{equation}
\partial_t u + \left( \frac{r}{\rho} \right) \partial_\rho u
= \frac{1-u}{r} \; ,
\end{equation}
and analogously for the $z$ boundaries.  We choose as initial
condition $u=1$, and evolve the above hyperbolic equation until we
reach a stationary state with some predetermined tolerance.  The idea
is that the wave-like equation above will propagate the residual away
through the boundaries, and will drive the system to a static solution
which corresponds to the solution of the original elliptic problem.

Spatial derivatives are approximated with standard centered second
order differences, while for the time integration we use a three-step
iterative Crank-Nicholson algorithm~\cite{Teukolsky00,Alcubierre99d}.
The resulting code is rather slow, particularly for high resolutions,
but has the advantage of being both very simple to write (and
debug), and also extremely robust.  A more sophisticated elliptic
solver should of course be used in order to find highly accurate
solutions in a short computational time (we certainly do not recommend
to use our quick-and-dirty ``wave-like'' algorithm for any kind of
production runs).

In order to deal with the problem of divisions by $\rho$ our
computational grid staggers the axis and introduces a ghost grid point
at position $\rho = - \Delta \rho / 2$.  To determine the value of $u$
at this ghost point we then simply impose the parity condition $u(-
\Delta \rho / 2) = u(\Delta \rho / 2)$.

Notice that we don't do anything special close to the punctures and
just take standard centered differences everywhere.  This clearly
affects the order of convergence close to the punctures (see numerical
examples below).

\subsubsection{Example I: Equal masses and opposite charges}

As a first example we will consider the case of two black holes with
equal masses $M_1=M_2=1$, and equal but opposite charges $Q_1 = - Q_2
= 1/2$. The punctures are located along the $z$ axis at positions
$z_1=-z_2=2$, and the boundaries extend to $\rho = 6$, $z = \pm 6$.

As initial guess we choose $u=1$ everywhere, and we then evolve
equation~\eqref{eq:wavelike} until the magnitude of its right hand
side is everywhere smaller than a fixed tolerance of
$\epsilon=10^{-8}$. This tolerance is chosen such that it is always
smaller than the truncation error at the resolutions considered here.

\begin{figure}
\epsfxsize=100mm
\centerline{\epsfbox{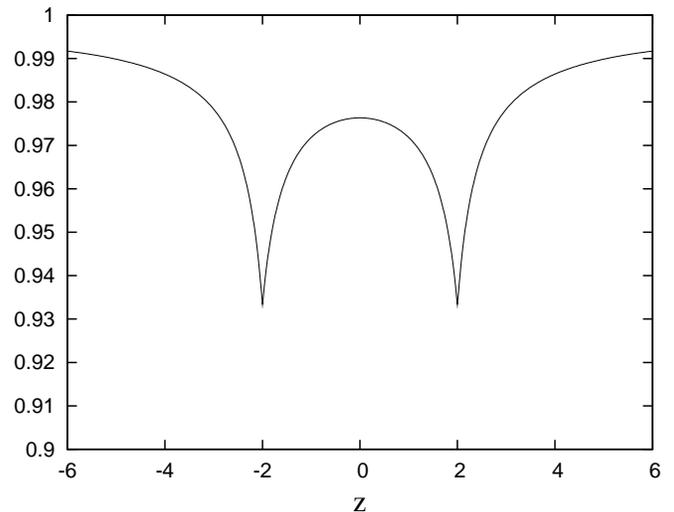}}
\caption{Numerical solution for the function $u$ along the $z$ axis
  for the case of equal masses and equal but opposite charges.}
\label{fig:u1}
\end{figure}

\begin{figure}
\epsfxsize=110mm \centerline{\epsfbox{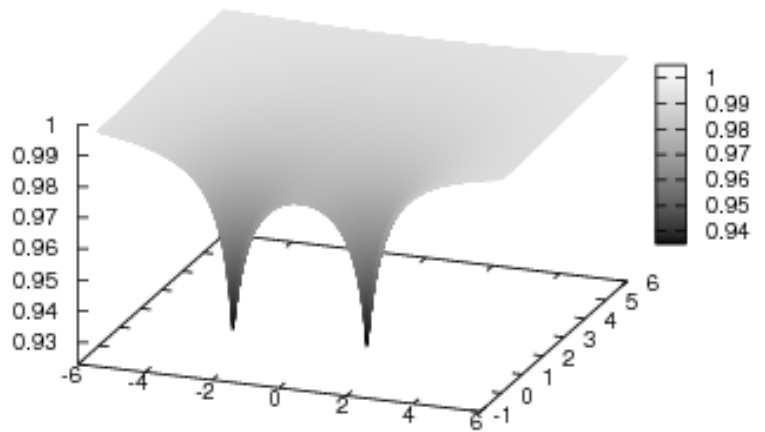}}
\caption{Height map in the $(\rho,z)$ plane of the numerical solution
  for the function $u$.}
\label{fig:u1_2d}
\end{figure}

Figure~\ref{fig:u1} shows the numerical solution for the function $u$
along the $z$ axis, for a resolution of $\Delta \rho = \Delta z =
0.0125$.  Notice how the function $u$ behaves as expected on both
punctures, with very evident kinks. One can also see that $u$ has
equatorial symmetry even though the charges have opposite signs.  Of
course, had we chosen both charges with the same sign the exact
solution would have been $u=1$ everywhere since in that case the
charge-to-mass ratios would have been equal.  It is also interesting
to note that the maximum deviation of the function $u$ from unity is
precisely at the punctures, and this maximum deviation is of only
about $7 \%$.  Figure~\ref{fig:u1_2d} shows the a height map of the
same solution in the $(\rho,z)$ plane.

Next we show a plot of the convergence in the Hamiltonian constraint
along the $z$ axis.  Since we are in fact solving numerically
precisely the Hamiltonian constraint, one would expect it to be
satisfied to the level of the tolerance in the elliptic solver.  This
is of course true, but in order to study convergence we are in fact
using a different expression for the Hamiltonian constraint.  We first
reconstruct the conformal factor $\psi$, and later evaluate
numerically to second order the Hamiltonian
constraint~\eqref{eq:hamconf} written as:
\begin{equation}
\hat{D}^2 \psi + \frac{1}{4 \psi^3} \:
\partial_m \varphi \: \partial^m \varphi = 0 \; .
\label{eq:hamconvergence}
\end{equation}
This last expression should not be expected to hold to the level of
the tolerance in the elliptic solver, but rather to the level of
numerical truncation error, which is much higher (we have chosen a
very small tolerance parameter in the elliptic solver precisely for
this reason).

\begin{figure}
\epsfxsize=100mm
\centerline{\epsfbox{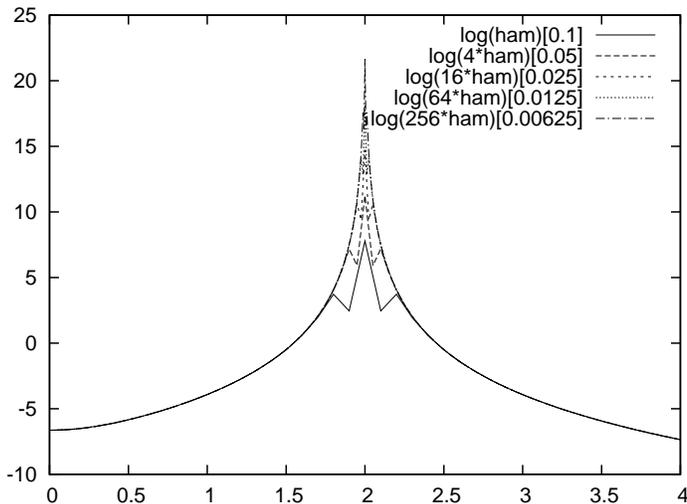}}
\caption{Convergence of the Hamiltonian constraint. We plot the
  logarithm of the absolute value of the Hamiltonian constraint
  evaluated at 5 different resolutions, with each resolution rescaled
  by the corresponding factor in order to show second order
  convergence.}
\label{fig:ham1}
\end{figure}

Figure~\ref{fig:ham1} shows the logarithm of the absolute value of the
Hamiltonian constraint evaluated using~\eqref{eq:hamconvergence} for
the five different resolutions $\Delta \rho = \Delta z =
0.1,0.05,0.025,0.0125,0.00625$, with each plot rescaled by the
corresponding factor expected for second order convergence:
$1,4,16,64,256$ (we in fact show only the region close to one of the
punctures as the situation is symmetric on the other puncture). Notice
how away from the puncture all plots lie on top of each other, showing
nice second order convergence.  Closer to the puncture, however, the
Hamiltonian constraint in fact increases with higher resolution, but
this loss of convergence is limited to the 2 or 3 grid points closest
to the puncture, so that the non-converging region keeps getting
smaller and smaller with higher resolution.  This should not be
surprising since in the conformal factor $\psi$ we have terms that
diverge as $1/r$ close to each puncture.

\subsubsection{Example II: Equal masses and one charge equal to zero}

\begin{figure}
\epsfxsize=100mm
\centerline{\epsfbox{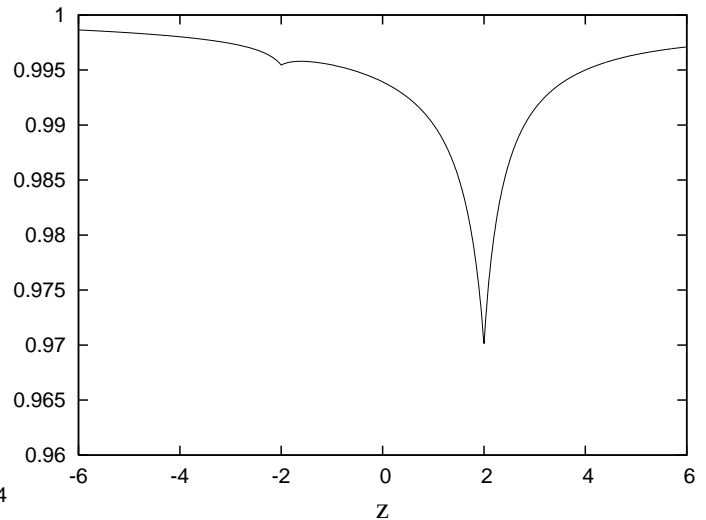}}
\caption{Similar to Fig.~\ref{fig:u1}, but for the case of equal
  masses and one charge equal to zero.}
\label{fig:u2}
\end{figure}

As a second example we will consider the case of two black holes with
equal masses $M_1=M_2=1$, and one charge set equal to zero $Q_1=1/2$,
$Q_2=0$.  As before, the punctures are located along the $z$ axis at
positions $z_1=-z_2=2$ and we use a resolution of $\Delta \rho =
\Delta z = 0.0125$.  The results are plotted in
Figure~\ref{fig:u2}. Notice that even though $Q_2=0$, the function $u$
still has a small kink at $z=-2$.

\subsubsection{Example III: Some more generic cases}

\begin{figure}
\epsfxsize=100mm \centerline{\epsfbox{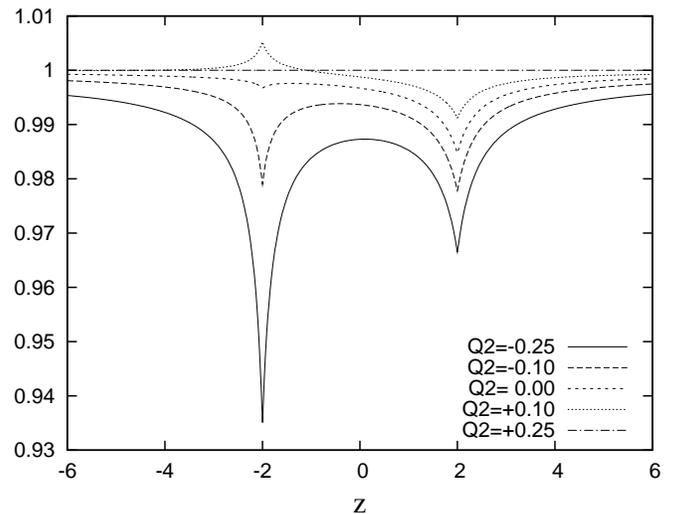}}
\caption{Similar to Fig.~\ref{fig:u1} for a more generic case in which
  the back hole at $z=2$ has $M_1=1$, $Q_1=1/2$, while for the black
  hole at $z=-2$ we take $M_2=1/2$ and a series of values for the
  charge: $Q_2=-0.25,-0.1,0,0.1,0.25$.}
\label{fig:u3}
\end{figure}

As a more generic case we set up two black holes with different
masses, $M_1=1$,$M_2=0.5$, located as before at the points
$z_1=-z_2=2$.  The first black hole has a charge of $Q_1 = 1/2$, while
for the charge of the second black hole we consider a series of
values: $Q_2=-0.25,-0.1,0.0,0.1,0.25$. The results are plotted in
Figure~\ref{fig:u3}.

What seems to happen as $Q_1>0$ is kept fixed and $Q_2$ changes is
that, for $Q_2<0$, the kinks in $u$ are point down at both punctures.
Then, as $Q_2$ approaches zero the kink at that puncture becomes much
smaller but still keeps pointing down. If $Q_2$ goes through zero and
becomes positive, then the kink at $z_2$ first disappears and later
starts growing again in the opposite direction.  As $Q_2$ keeps
increasing towards a value where both charge-to-mass ratios are equal
both kinks become smaller with opposite directions, until the function
$u$ becomes 1 everywhere.

\begin{figure}
\epsfxsize=100mm \centerline{\epsfbox{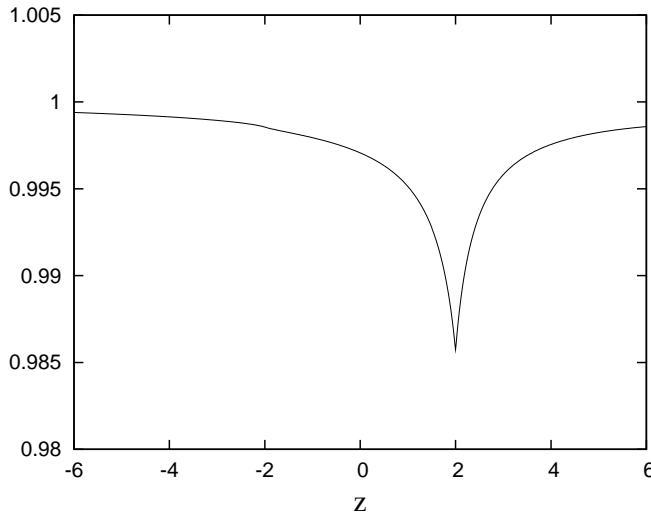}}
\caption{Solution for $u$ in the case when
  $M_1=1,M_2=0.5,Q_1=0.5,Q_2=0.01475$. Notice that the kink at $z=-2$
  is essentially gone.}
\label{fig:u4}
\end{figure}

By trial and error we have in fact found a situation for which the
kink at $z=-2$ seems to essentially disappear.  Figure~\ref{fig:u4}
shows the numerical solution for $u$ in the case when
$M_1=1,M_2=0.5,Q_1=0.5,Q_2=0.01475$.

The special value of $Q_2$ for which the kink disappears can in fact
be estimated analytically. When we examined the behaviour of the
function $u$ near the punctures in Section~\ref{sec:numericaldata}
above, we mentioned the fact the source term of
equation~\eqref{eq:hamconf4}, which we called $T_2$, behaves as
$1/r^4$ near each puncture.  Being somewhat more precise one can show
that for the case of two black holes this term behaves close to the
puncture at $\vec{r}_2$ as
\begin{eqnarray}
T_2 &=& \frac{1}{4 r^4} \: \left( \frac{Q_1 M_2}{d} - \frac{Q_2 M_1}{d}
- 2 Q_2 \Delta u \right) \nonumber \\
&\times& \left( \frac{Q_1 M_2}{d} - \frac{Q_2 M_1}{d} - 4 Q_2 
- 2 Q_2 \Delta u \right) \nonumber \\
&+& {\cal O} (1/r^3) \; ,
\end{eqnarray}
with $r=|\vec{r} - \vec{r}_2|$, and where $d$ is the distance between
the punctures and $\Delta u = u(\vec{r}_2)-1$.  Notice that there are
two ways in which the term $T_2$ can vanish, corresponding to a
solution $u$ that is regular at the puncture.  One possibility is to
have $u=1$, so that $\Delta u=0$, and equal charge-to-mass ratios so
that $Q_1 M_2 = Q_2 M_1$, in which case the first term above vanishes.
This is clearly the exact solution we already described. But a second
possibility is to assume that $u$ is very close to 1 at the puncture
so that $\Delta u \ll 1$, and to ask for $Q_2 \sim Q_1 M_2 / (M_1 +
4 d)$, in which case the second term above will almost vanish ($T_2$
will in fact not vanish exactly for this value of $Q_2$ since at the
puncture $\Delta u$ is not exactly zero, but one can expect $T_2$ to
vanish for a value of $Q_2$ that is very close to this one).  For the
values in our example we have $M_1=1$, $M_2=0.5$, $Q_1=0.5$ and $d=4$,
so that $Q_2 \sim 1/68 \sim 0.0147$, which is remarkably close to the
empirical value found above.

%%%%%%%%%%%%%%%%%%%%%%%
%%%   CONCLUSIONS   %%%
%%%%%%%%%%%%%%%%%%%%%%%

\section{Conclusions}
\label{sec:conclusions}

We have considered the Einstein-Maxwell system having two goals in
mind.  The first goal consisted in recasting the covariant Maxwell
equations in a curved spacetime as an initial value problem along the
lines of the usual 3+1 formalism of general relativity, choosing the
magnetic and electric fields as independent variables.  This led to a
set of two constraint equations for the electric and magnetic fields,
plus a set of two evolution equations for those fields. The evolution
equations are hyperbolic with the propagation speeds depending on the
lapse and shift. 

The second goal was to construct initial data satisfying the
gravitational and electromagnetic constraints which represent
momentarily static charged black holes.  In order to achieve this goal
we assumed a moment of time symmetry with vanishing extrinsic
curvature and magnetic field.  For the case of two black holes, this
initial data will serve to analyze a head-on collision. We found that
for black holes having an equal charge-to-mass ratio it is possible to
find an analytic solution for the constraints. However, when this
condition is dropped, we present instead numerical solutions. We
studied the behavior of such solutions with several values of the free
parameters ({\em i.e.} different masses and charges), showing extremal
cases.

A much more realistic initial data will consist in relaxing the moment
of time symmetry condition ({\em i.e.} abandon the initial condition
$K_{ij}=0$), but keeping a null initial magnetic field and analyze the
analogous of the Bowen-York initial data. In this case, one will need
to solve a much more complex Hamiltonian constraint for the conformal
factor. An elliptic solver will be required {\em a fortiori}\/ to solve
this constraint. This is an issue that is worth analyzing in the
future.

Using the initial data presented here we plan in a near future to
study the evolution of the Einstein-Maxwell system generated by the
collision of two charged black holes, and analyze the emission of both
gravitational and electromagnetic radiation. We believe that the
interplay between gravity and electromagnetism will give rise to very
interesting and unusual dynamics due to the possibility of repulsion
between charges. This is an issue that has not been scrutinized thus
far in numerical relativity.

%%%%%%%%%%%%%%%%%%%%%%%%%%%
%%%   ACKNOWLEDGMENTS   %%%
%%%%%%%%%%%%%%%%%%%%%%%%%%%

\begin{acknowledgments}

This work was supported in part by Direcci\'on General de Estudios de
Posgrado (DGEP-UNAM), by CONACyT through grant 82787, and by
DGAPA-UNAM through grant IN113907. JCD also acknowledges CONACyT support.

\end{acknowledgments}

%%%%%%%%%%%%%%%%%%%%%%
%%%   APPENDIX A   %%%
%%%%%%%%%%%%%%%%%%%%%%

\appendix

\section*{Appendix}
\label{sec:appendix}

For the sake of clarity, we will give here some details about our
derivation of the 3+1 Maxwell equations. We start with
Eq.~\eqref{Maxcov1}. The projection of this equation onto $n_a$ reads:
\begin{equation}
n_b \nabla_a F^{ab} = 4\pi \rho \; ,
\end{equation}
where one must remember $\rho:= - n_b j^b$. At this point there are
several possible ways to proceed. Let us first write
\begin{eqnarray}
n_b \nabla_a F^{ab} &=& \nabla_a \left(n_b F^{ab}\right)
- F^{ab} \nabla_a n_b \nonumber \\
&=& \nabla_a E^a -  F^{ab} \nabla_a n_b \; ,
\label{proyMaxcov1}
\end{eqnarray}
where in the last step we used Eq.~\eqref{Electric}.

Now, concerning the term $\nabla_a E^a$, one way to relate it with 3+1
quantities is by using the following identity for any 4-vector field
$V^a$
\begin{eqnarray}
\nabla_a V^a &=& \frac{1}{\sqrt{-g}} \; \partial_a \left( \sqrt{-g} V^a \right)
\nonumber \\
&=& \frac{1}{N\sqrt{h}} \; \partial_a \left(N \sqrt{h} V^a \right) \; ,
\end{eqnarray}
where we have used the fact that $g= - N^2 h$. We then have
\begin{equation}
\nabla_a V^a = V^a \partial_a {\rm ln} N + \frac{1}{\sqrt{h}}
\; \partial_a\left(\sqrt{h} V^a\right) \; .
\end{equation}
For the particular case where $V^a= E^a$, it turns out that $E^a
\partial_a {\rm ln} N = E^b h^{\,\,a}_{b} \nabla_a {\rm ln} N = E^a
D_a {\rm ln} N$, where the fact that $E^a\equiv E^b h^{\,\,a}_{b} $ is
a consequence of the fact that $E^a$ is a 3-vector. Moreover, since
$E^t\equiv 0$ the term $ (1 / \sqrt{h}) \; \partial_a (\sqrt{h} E^a
)$ reduces to the 3-divergence $D_a E^a$. We then have
\begin{equation}
\label{nablaE}
\nabla_a E^a = D_a E^a + E_a a^a \; ,
\end{equation}
where we used the following identity for the acceleration of the
Eulerian observer $a_b:= n^a \nabla_a n_b \equiv D_b {\rm ln} N$.  We
point out that one could have obtained the same result from the
definition of the 3-covariant divergence:
\begin{eqnarray}
D_a E^a &=& h^{\,\,c}_{a} h^{a}_{\,\,b} \nabla_c E^b
= h^{\,\,c}_{b} \nabla_c E^b = \nabla_b E^b + n^cn_b \nabla_c E^b
\nonumber \\
&=& \nabla_b E^b + n^c \nabla_c \left(n_b E^b\right)
- n^c E^b \nabla_c n_b \nonumber \\
&=& \nabla_b E^b -  E^b a_b \; .
\end{eqnarray}
where we used the fact that $n_b E^b\equiv 0$ since $E^a$ is by
definition orthogonal to $n^a$. From this equation one obtains the
same identity~\eqref{nablaE}.  Using this result
Eq.~\eqref{proyMaxcov1} now reads
\begin{equation}
\label{proyMaxcov2}
n_b \nabla_a F^{ab} = D_a E^a + E_a a^a -F^{ab} \nabla_a n_b \,\,\,.
\end{equation}
We now need to prove that, in fact, $F^{ab} \nabla_a n_b= E_a
a^a$. For this we use Eq.~\eqref{Fdecomp} to obtain
\begin{equation}
F^{ab} \nabla_a n_b = \: ^{(3)}\! F^{ab} \nabla_a n_b
+ n^a E^b \nabla_a n_b - E^a n^b \nabla_a n_b .
\end{equation}
The second term in the last equation is precisely $n^a E^b \nabla_a
n_b= E^b a_b$, while the third vanishes identically since $n^b n_b
=-1$, and therefore $\nabla_a \left(n^b n_b\right)= 2 n^b \nabla_a n_b
\equiv 0$. Finally, we show that the first term vanishes by using the
definition of the extrinsic curvature, Eq.~\eqref{K_ab}:
\begin{equation}
^{(3)}\! F^{ab} K_{ab}
= -\, ^{(3)} \! F^{ab} h^{\,\,c}_{a} h^{\,\,d}_{b} \nabla_c n_d
= -\, ^{(3)} \! F^{cd} \nabla_c n_d .
\end{equation}
This term vanishes identically because $^{(3)}\:\! F^{ab} $ is
antisymmetric while $K_{ab}$ is symmetric.  We have thus proved that
$F^{ab} \nabla_a n_b = E_a a^a$. Using this result in
Eq.~\eqref{proyMaxcov2} allows us to conclude that
\begin{equation}
n_b \nabla_a F^{ab} = D_a E^a = 4\pi \rho \; .
\end{equation}

Following an analogous procedure, one can now also obtain
Eq~\eqref{Max4} by projecting Eq.~\eqref{Maxcov2}, except that one now
uses Eq.~\eqref{FdcompDual} instead of \eqref{Fdecomp}. In this case,
there is no magnetic charge.

\bigskip

We proceed now to obtain Eq.~\eqref{evE} by projecting \eqref{Maxcov1}
onto $\Sigma_t$ using $h_{\,\,\,b}^{a}$:
\begin{equation}
h_{\,\,\,b}^{a}\nabla_c F^{cb} =  -4\pi \,^{(3)}\,\!j^a \; .
\end{equation}

Using Eq. (\ref{Fdecomp}) and expanding we obtain
\begin{eqnarray}
&& \hspace{-10mm} h_{\,\,\,b}^{a} \nabla_c \, ^{(3)}\,\!F^{cb}
- E^c h_{\,\,\,b}^{a} \nabla_c n^b + E^a \nabla_c n^c
+ h_{\,\,\,b}^{a} n^c \nabla_c E^b  
\nonumber \\
\label{evEp1}
&=& h_{\,\,\,b}^{a} \nabla_c \, ^{(3)}\,\!F^{cb} + E^b K_{b}^{\,\,\,a}
- E^a K + h_{\,\,\,b}^{a} n^c \nabla_c E^b \nonumber \\
&=& -4 \pi \,^{(3)}\,\!j^a \ .
\end{eqnarray}
where we used $ E^c h_{\,\,\,b}^{a} \nabla_c n^b= E^d h_{d}^{\,\,\,c}
h_{\,\,\,b}^{a} \nabla_c n^b = - E^d K_{d}^{\,\,\,a}$, and $K=
-\nabla_c n^c$ [cf. Eqs. \eqref{K_abexp} and \eqref{TraceK}].

Now, from the definition of the 3-covariant derivative applied to a
3-tensor field we have [cf. Eq. \eqref{DT}]
\begin{eqnarray}
D_c \: ^{(3)} \!F^{cb} &=& h_{c}^{\,\,\,d} h_{\,\,\,e}^{c}
h_{\,\,\,f}^{b} \nabla_d \, ^{(3)}\,\!F^{ef}
= h_{\,\,\,e}^{d} h_{\,\,\,f}^{b} \nabla_d \, ^{(3)}\,\!F^{ef} \nonumber \\
&=& h_{\,\,\,f}^{b} \nabla_d \, ^{(3)}\,\!F^{df} - \, ^{(3)}\,\!F^{eb}
\nabla_d \left(n_e n^d\right) \nonumber \\
&=& h_{\,\,\,f}^{b} \nabla_d \, ^{(3)}\,\!F^{df} - \, ^{(3)}\,\!F^{eb} a_e \; .
\end{eqnarray}
In this way Eq. \eqref{evEp1} becomes
\begin{eqnarray}
h_{\,\,\,b}^{a} n^c \nabla_c E^b + D_c \, ^{(3)}\,\!F^{ca}
+ \, ^{(3)}\,\!F^{ba} a_b \nonumber \\
+  E^b K_{b}^{\,\,\,a} - E^a K
= -4\pi \,^{(3)}\,\!j^a \; .
\end{eqnarray}
The first term above can be rewritten using the following identities
\begin{eqnarray}
h_{\,\,\,b}^{a} {\cal L}_{\mathbf{n}} E^b
&=& h_{\,\,\,b}^{a}\left( n^c\nabla_c E^b - E^c \nabla_c n^b\right) \nonumber \\
&=& h_{\,\,\,b}^{a} n^c\nabla_c E^b + K^a_{\,\,\,c} E^c \; .
\end{eqnarray}
We then conclude that
\begin{eqnarray}
\label{evEp2}
h_{\,\,\,b}^{a} {\cal L}_{\mathbf{n}} E^b + D_c ^{(3)} \!F^{ca}
+ ^{(3)} \!F^{ba} a_b \hspace{10mm} \nonumber \\
- E^a K = - 4 \pi \:^{(3)} \!j^a \; .
\end{eqnarray}

Finally, using the fact that $^{(3)}\,\!F^{ca}$ is antisymmetric,
together with Eqs. \eqref{3FtoB} and \eqref{Levi-Civita-flatup}, one
can write
\begin{eqnarray}
D_c \, ^{(3)}\,\!F^{ca}
&=& \frac{1}{\sqrt{h}} \: \partial_i\left(\sqrt{h} \, ^{(3)}\,\!F^{ia} \right)
= \frac{1}{\sqrt{h}} \: \partial_b\left( \epsilon^{bac}_F B_c \right) \nonumber \\
&=& \frac{\epsilon^{bac}_F}{\sqrt{h}} \: \partial_b  B_c
= - \,^{(3)} \,\!\epsilon^{abc} \partial_b  B_c \; .
\end{eqnarray}
Equation (\ref{evEp2}) then reads 
\begin{eqnarray}
\label{evEp3}
h_{\,\,\,b}^{a} {\cal L}_{\mathbf{n}} E^b
- \,^{(3)} \,\!\epsilon^{abc} \partial_b  B_c
+ \,^{(3)} \,\!\epsilon^{abc} B_b a_c \hspace{10mm} \nonumber \\
 - E^a K = - 4 \pi \,^{(3)}\,\!j^a \; ,
\end{eqnarray}
where we used again Eq. \eqref{3FtoB} to rewrite the third term.
Taking the notation given by Eqs. \eqref{rotB} and \eqref{Bxa}, one
finally recovers Eq. \eqref{DynE}.

In order to find Eq. \eqref{evE} from Eq. \eqref{evEp3}, one has to
remember that the Lie derivative can be written in terms of ordinary
derivatives as follows
\begin{equation}
h_{\,\,\,b}^{a} {\cal L}_{\mathbf{n}} E^b
= h_{\,\,\,b}^{a}\left( n^c\partial_c E^b - E^c \partial_c n^b\right) \; .
\end{equation}
Now, since we are only interesting in the spatial components of
Eq. \eqref{evEp3}, using \eqref{proj} together with $n_a=
(-N,0,0,0)$, $n^a = (1/N,N^i/N)$ one can easily show that
\begin{equation}
h_{\,\,\,a}^{i} {\cal L}_{\mathbf{n}} E^a
= \frac{1}{N} \: \partial_t E^i + \frac{N^j}{N} \; \partial_j E^i
- \frac{E^j}{N} \: \partial_j N^i \; .
\end{equation}
Substituting this last result into the spatial components of
\eqref{evEp3} leads directly to Eq. \eqref{evE}.

In a similar fashion one can obtain Eq. \eqref{evB} by projecting
\eqref{Maxcov2} onto $\Sigma_t$. In fact, this just amounts to using
the duality relations $E_a \rightarrow B_a$ and $B_a \rightarrow -E_a$
in~\eqref{evE} to obtain Eq. \eqref{evB}.

\bigskip

Finally, we proceed to derive Eq. \eqref{chargecons}. This can be
easily done by using the orthogonal decomposition of the
electric-current 4-vector [cf. Eq. \eqref{ortdecomw}]:
\begin{equation}
j^a = \,^{(3)}j^a + \rho n^a \,\,\,\,,
\end{equation}
being $\rho:= - n_a j^a$ the charge density measured by the Eulerian
observer, and $\,^{(3)}j^a:= h_{\,\,\,c}^{a} j^c$.  This
decomposition, when inserted into Eq. \eqref{CQ} and making use of
\eqref{TraceK}, leads directly to
\begin{equation}
n^a\nabla_a \rho - \rho K + \nabla_a \,^{(3)}j^a = 0 \; .
\end{equation}
Using now Eq. \eqref{nablaE} for $\,^{(3)}j^a$ instead of $E^a$ (the
expression is valid for any 3-vector), one obtains
\begin{equation}
{\cal L}_{\mathbf{n}} \rho + D_a \,^{(3)}j^a
+ \,^{(3)}j^a a_a - \rho K  = 0 \; ,
\end{equation}
where
\begin{equation}
{\cal L}_{\mathbf{n}} \rho \equiv n^a\nabla_a \rho
= 1/N \left(\partial_t \rho + N^j\partial_j \rho \right) \; .
\end{equation}

%%%%%%%%%%%%%%%%%%%%%%
%%%   REFERENCES   %%%
%%%%%%%%%%%%%%%%%%%%%%

\bibliographystyle{bibtex/prsty}
\bibliography{bibtex/referencias}

%%%%%%%%%%%%%%%
%%%   END   %%%
%%%%%%%%%%%%%%%

\end{document}